\documentclass[journal]{IEEEtran}
\IEEEoverridecommandlockouts
\usepackage{lineno}
\usepackage{hyperref}
\usepackage{cite}
\usepackage{amsmath,amssymb,amsfonts,cases} 
\usepackage{amsmath}
\usepackage{amsthm}

\usepackage{graphicx}
\usepackage{textcomp}
\usepackage{xcolor}
\usepackage{graphicx}
\usepackage{float}
\usepackage{subfigure}
\usepackage{amsmath,mleftright}
\usepackage{amsfonts,amssymb}
\usepackage{mathrsfs}
\usepackage{mathtools}
\usepackage{algorithm}
\usepackage{algorithmic}
\usepackage{bm}
\usepackage{multirow}
\usepackage{array}
\usepackage{amssymb}
\usepackage{amsmath}
\usepackage{cite}
\usepackage{url}
\usepackage{xcolor}
\usepackage{cite,graphicx,amsmath,amssymb}
\usepackage{subfigure}
\usepackage{fancyhdr}
\usepackage{mdwmath}
\usepackage{mdwtab}
\usepackage{caption}
\usepackage{amsthm}
\usepackage{setspace}
\usepackage{bm}
\usepackage{mathtools}
\usepackage{dsfont}
\usepackage{bbm}
\usepackage{framed}
\newtheorem{remark}{Remark}
\newtheorem{theorem}{Theorem}

\newtheorem{lemma}{Lemma}

\newtheorem{corollary}{Corollary}

\makeatletter
\newcommand{\biggg}{\bBigg@{3}}
\newcommand{\Biggg}{\bBigg@{3.5}}
\makeatother
\makeatletter
\renewcommand{\maketag@@@}[1]{\hbox{\m@th\normalsize\normalfont#1}}%
\makeatother
\def\BibTeX{{\rm B\kern-.05em{\sc i\kern-.025em b}\kern-.08em
    T\kern-.1667em\lower.7ex\hbox{E}\kern-.125emX}}
    \expandafter\def\expandafter\normalsize\expandafter{%
    \normalsize%
    \setlength\abovedisplayskip{4pt}%
    \setlength\belowdisplayskip{4pt}%
    \setlength\abovedisplayshortskip{2pt}%
    \setlength\belowdisplayshortskip{2pt}%
}
\begin{document}
\title{Uplink and Downlink Communications in Segmented Waveguide-Enabled Pinching-Antenna Systems (SWANs)}
\author{Chongjun~Ouyang, Hao~Jiang, Zhaolin~Wang, Yuanwei~Liu, and Zhiguo~Ding\vspace{-10pt}
\thanks{C. Ouyang and H. Jiang are with the School of Electronic Engineering and Computer Science, Queen Mary University of London, London, E1 4NS, U.K. (e-mail: \{c.ouyang, hao.jiang\}@qmul.ac.uk).}
\thanks{Z. Wang and Y. Liu are with the Department of Electrical and Electronic Engineering, The University of Hong Kong, Hong Kong (email: \{zhaolin.wang, yuanwei\}@hku.hk).}
\thanks{Z. Ding is with the School of Electrical and Electronic Engineering, The University of Manchester, Manchester, M13 9PL, U.K., and also with the Department of Electrical Engineering and Computer Science, Khalifa University, Abu Dhabi, UAE (e-mail: zhiguo.ding@manchester.ac.uk).}}
\maketitle
\begin{abstract}
A segmented waveguide-enabled pinching-antenna system (SWAN) is proposed, in which a segmented waveguide composed of multiple short dielectric waveguide segments is employed to radiate or receive signals through the pinching antennas (PAs) deployed on each segment. Based on this architecture, three practical operating protocols are proposed: segment selection (SS), segment aggregation (SA), and segment multiplexing (SM). For uplink SWAN communications, where one PA is activated per segment, the segmented structure eliminates the inter-antenna radiation effect, i.e., signals captured by one PA may re-radiate through other PAs along the same waveguide. This yields a tractable and physically consistent uplink signal model for a multi-PA pinching-antenna system (PASS), which has not been established for conventional PASS using a single long waveguide. Building on this model, PA placement algorithms are proposed to maximize the uplink signal-to-noise ratio (SNR). Closed-form expressions for the received SNR under the three protocols are derived, and the corresponding scaling laws with respect to the number of segments are analyzed. It is proven that the segmented architecture reduces both the average PA-to-user distance and the PA-to-feed distance, thereby mitigating both large-scale path loss and in-waveguide propagation loss. These results are extended to downlink SWAN communications, where multiple PAs are activated per segment, and PA placement methods are proposed to maximize the downlink received SNR under the three protocols. Numerical results demonstrate that: \romannumeral1) among the three protocols, SM achieves the best performance, followed by SA and then SS; and \romannumeral2) for all protocols, the proposed SWAN achieves a higher SNR than conventional PASS with a single long waveguide in both uplink and downlink scenarios.
\end{abstract}
\section{Introduction}
Over the past three decades, multiple-input multiple-output (MIMO) technology has fundamentally reshaped wireless communications \cite{heath2018foundations}. The field is now entering a new phase in which research increasingly emphasizes \emph{reconfigurable antennas}. By dynamically adjusting electromagnetic (EM) properties such as polarization, operating frequency, and radiation pattern \cite{heath2025tri}, these antennas enable flexible \emph{EM beamforming}. This capability facilitates efficient channel manipulation and improves system performance without substantially increasing radio-frequency (RF) complexity or power usage \cite{heath2025tri,castellanos2023energy}.

Typical examples of reconfigurable antennas include fluid antennas \cite{wong2020fluid} and movable antennas \cite{zhu2024movable}, which adjust their physical positions to exploit spatial diversity and mitigate small-scale fading. While these technologies offer promising performance improvements, they remain limited in addressing \emph{large-scale path loss}. Their reconfiguration capability is typically confined to apertures spanning only a few to several tens of wavelengths, which is insufficient to overcome large-scale path loss. Moreover, once these systems are deployed, modifying their antenna configurations, such as adding or removing antennas, incurs substantial cost and complexity. 

\begin{figure}[!t]
\centering
\includegraphics[width=0.35\textwidth]{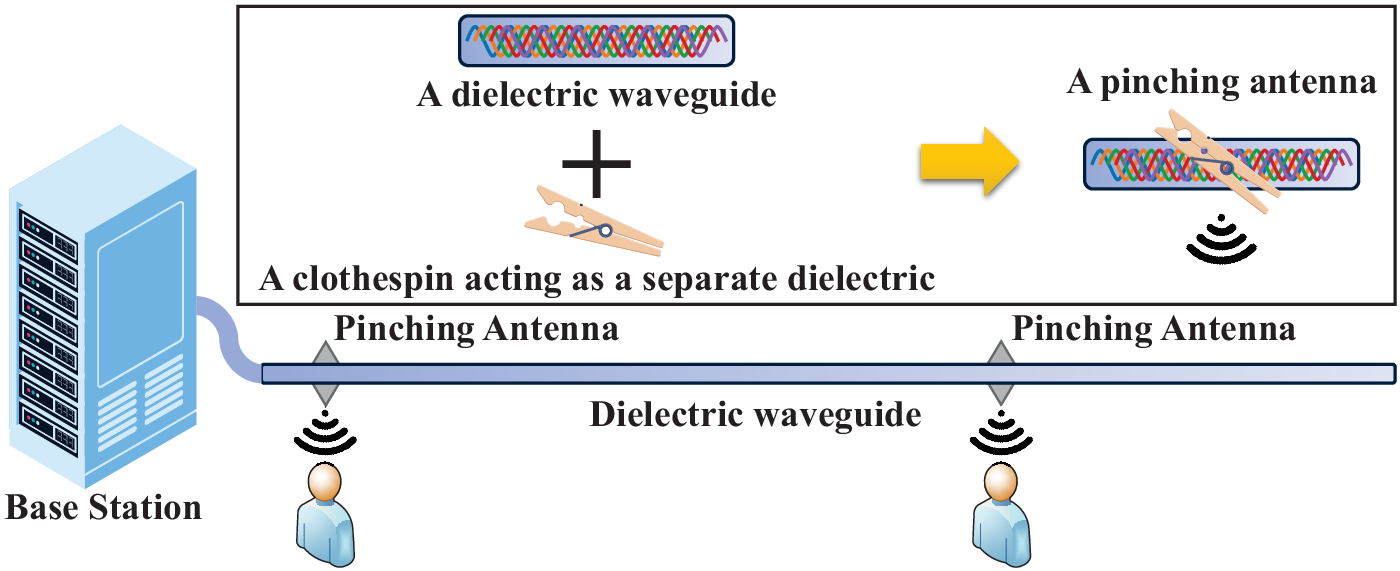}
\caption{Illustration of PASS.}
\label{Figure_PASS}
\vspace{-15pt}
\end{figure}

\subsection{Pinching Antennas}
To address the aforementioned limitations, NTT DOCOMO recently introduced the \emph{Pinching-Antenna SyStem (PASS)} \cite{suzuki2022pinching}, a novel form of reconfigurable-antenna technology. PASS employs a dielectric waveguide \cite{pozar2021microwave} as the transmission medium, where EM waves are radiated by attaching small dielectric particles at designated locations along the waveguide \cite{suzuki2022pinching}. These dielectric elements are typically mounted on the tips of plastic clips, forming what are referred to as pinching antennas (PAs). Each PA can be independently activated or deactivated at any point along the waveguide, enabling dynamic reconfiguration of the antenna array \cite{suzuki2022pinching}. This concept resembles attaching or removing clothespins on a clothesline \cite{yang2025pinching,liu2025pinching}, as shown in {\figurename} {\ref{Figure_PASS}}. Unlike conventional reconfigurable antennas, the waveguide in PASS can be extended to arbitrary lengths, which allows antennas to be positioned in close proximity to users and establish strong and stable line-of-sight (LoS) links, thereby reducing large-scale path loss. Moreover, PASS enables low-cost and easy deployment, as antenna reconfiguration requires only the addition or removal of dielectric materials.

Due to these distinctive characteristics, PASS has attracted significant attention in the wireless communications community. The average user rate achieved by PAs serving mobile users was first analyzed in \cite{ding2024flexible}. Subsequent studies investigated outage probability \cite{tyrovolas2025performance,ding2025blockage} and array gain \cite{ouyang2025array}. Collectively, these works demonstrated that PASS effectively mitigates large-scale path loss and outperforms both traditional fixed-antenna systems and existing fluid/movable-antenna technologies. Building on this foundation, several optimization algorithms have been proposed to determine efficient PA placements along the waveguide, so as to improve overall communication performance \cite{xu2024rate,wang2024antenna,tegos2024minimum,zeng2025energy,papanikolaou2025resolving}. Besides communication-focused studies, researchers have also begun exploring PASS in wireless sensing \cite{wang2025wireless} and integrated sensing and communications (ISAC) \cite{ouyang2025isac,ouyang2025rate2}. These works further highlight the flexibility of PASS and its ability to mitigate large-scale path loss while enhancing wireless capacity. A comprehensive overview of recent research progress can be found in the tutorial on PASS presented in \cite{liu2025pinchingtutorial}.
\subsection{Motivation and Contributions}
Despite significant research progress in PASS, existing architectures face three key challenges.
\subsubsection{Uplink Signal Model}
The downlink signal model of PASS has been clearly characterized, where signals injected at the waveguide feed point radiate passively from the activated PAs based on the EM coupling model proposed in \cite{wang2025modeling,liu2025pinchingtutorial}. In contrast, formulating a tractable and physically consistent uplink signal model is much more challenging. This difficulty arises because PAs can passively receive EM signals from free space into the waveguide. When multiple PAs are deployed along the same waveguide, signals captured by one PA may re-radiate through other PAs as they propagate toward the feed point. This \emph{inter-antenna radiation (IAR)} effect complicates the uplink analysis and makes the signal model mathematically intractable. As a result, most existing uplink studies restrict attention to single-PA deployments, thereby avoiding IAR, or they neglect IAR altogether in multi-PA cases, which leads to oversimplified models \cite{liu2025pinchingtutorial}. At present, no tractable and physically consistent uplink signal model for multi-PA PASS exists, leaving research in this direction stalled.
\subsubsection{In-Waveguide Propagation Loss}
Another challenge arises when a single long waveguide is deployed to cover a large region. While flexible PA placement can reduce the average user-to-antenna distance and thereby alleviate large-scale path loss, a long waveguide simultaneously increases the propagation distance between PAs and the feed point. This, in turn, increases in-waveguide propagation loss. Such loss is negligible for small or moderate waveguide lengths, but becomes a major limiting factor in large-scale deployments. The numerical results in this work demonstrate that in-waveguide loss can significantly degrade system performance when long waveguides are employed.
\subsubsection{Waveguide Maintainability}
Finally, long waveguide deployments present practical challenges in terms of reliability and maintenance. If a waveguide is damaged at a particular point, fault detection may be difficult, and replacing the entire waveguide can be both costly and complex. This lack of robustness reduces the practical feasibility of long-waveguide PASS deployments, as they are fragile and difficult to repair once installed.
\subsubsection*{Our Proposed Solution}
\begin{figure}[!t]
\centering
    \subfigure[System setup.]
    {
        \includegraphics[width=0.45\textwidth]{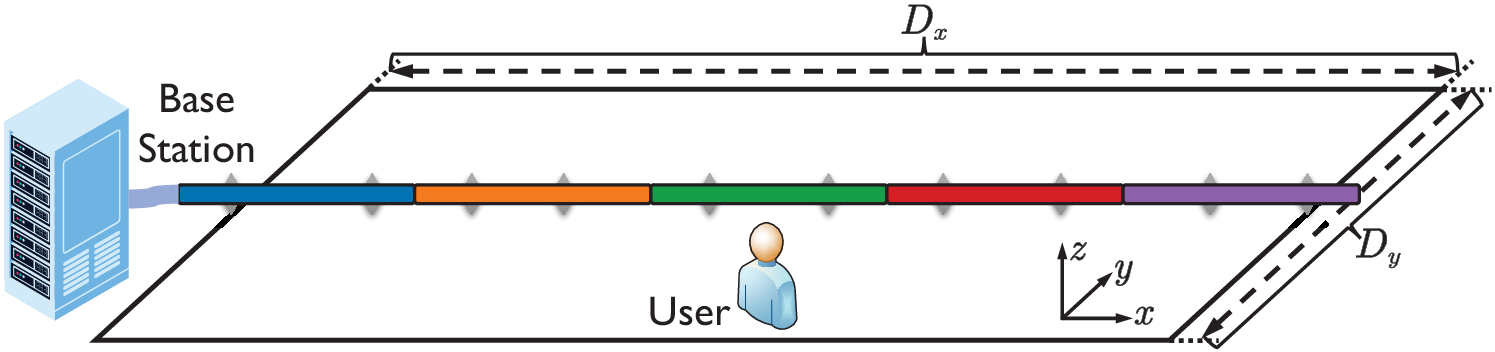}
	   \label{Figure: PAS_System_Model1}
    }
   \subfigure[Segmented waveguide.]
    {
        \includegraphics[width=0.45\textwidth]{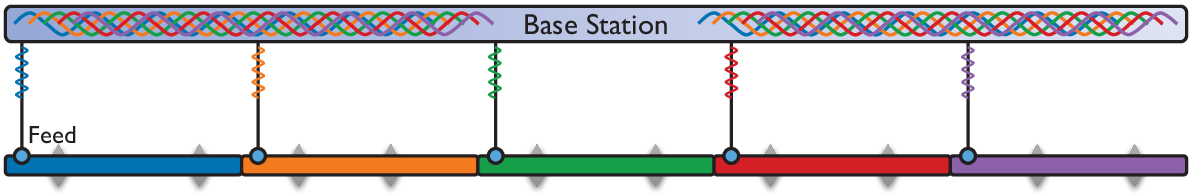}
	   \label{Figure: PAS_System_Model2}
    }
\caption{Illustration of the proposed SWAN architecture.}
\label{Figure: PAS_System_Model}
\vspace{-15pt}
\end{figure}

To address the aforementioned challenges, we propose the \emph{Segmented Waveguide-enabled pinching-ANtenna system (SWAN)}. The SWAN architecture employs multiple short dielectric waveguide segments arranged end-to-end. Unlike a single long waveguide, these segments are \emph{not physically interconnected}. Instead, each segment has its own feed point, through which signals are injected into or extracted from the waveguide and then relayed to the base station (BS) via wired connections such as optical fiber or high-quality coaxial cables, as illustrated in {\figurename} {{\ref{Figure: PAS_System_Model}}}. 

The proposed SWAN resolves the limitations of conventional PASS in three aspects. First, by activating only a single PA within each segment, the SWAN enables the realization of a multi-PA uplink PASS without IAR effects. This approach allows the system to achieve multi-PA array gain while maintaining a tractable uplink signal model. Second, because each segment is much shorter than the overall service region, the SWAN ensures both a reduced average PA-to-user distance and a short PA-to-feed distance. This dual benefit mitigates large-scale path loss as well as in-waveguide propagation loss. Third, the segmented design enhances maintainability, since failures can be localized and addressed at the segment level rather than requiring replacement of the entire waveguide. Short segments are easier to detect, repair, and replace compared to a single long waveguide. Finally, although long dielectric waveguides can be fabricated in principle, existing PASS prototypes developed by NTT DOCOMO employ short waveguides with lengths ranging from $0.4$ m to $1.2$ m \cite{yamamoto2021pinching,reishi2022pinching,junya2024pinching,junya2025pinching}. Thus, SWAN can be readily implemented by leveraging existing PASS hardware with only minimal architectural modifications.

To exploit the full potential of SWAN, we propose practical protocols for its operation and analyze its performance in both multi-PA uplink and downlink communications. The main contributions of this work are summarized as follows.
\begin{itemize}
  \item We propose the architecture of SWAN and establish its uplink and downlink signal models under multi-PA deployment. To process the signals radiated or extracted from different segments at the baseband, we introduce three practical operating protocols, namely \emph{segment selection (SS)}, \emph{segment aggregation (SA)}, and \emph{segments multiplexing (SM)}, and discuss their respective benefits and drawbacks.
  \item We analyze the performance of the SWAN-based uplink communications by considering the activation of one PA per segment. For each protocol, we develop PA placement algorithms that optimize the uplink received SNR. We further derive closed-form expressions for the SNR under all three protocols and characterize the SNR scaling law with respect to (w.r.t.) the number of segments. Our theoretical analysis proves that the SWAN yields significantly lower in-waveguide propagation loss than conventional uplink PASS. In addition, we show that the SWAN achieves higher array gain than conventional uplink PASS due to the simultaneous activation of multiple PAs.
  \item We extend the study to downlink communications with the SWAN, where multiple PAs are activated within each segment. For each protocol, we propose PA placement methods that optimize PA positions within segments to maximize the received SNR. We also provide a theoretical analysis showing that downlink SWAN reduces in-waveguide propagation loss and achieves superior performance compared to conventional downlink PASS.
  \item We present numerical results to validate the theoretical analysis and demonstrate that: \romannumeral1) in both uplink and downlink, SM achieves the performance upper bound of the SWAN, followed by SA and then SS; \romannumeral2) the SWAN outperforms conventional PASS in both uplink and downlink, with the performance gain increasing with the side length of the service region; and \romannumeral3) in SA, activating all segments is not always optimal, as there exists a segment number that maximizes system performance.
\end{itemize}

The remainder of this paper is organized as follows. Section {\ref{Section: System Model}} introduces the system model of the SWAN. Section \ref{Section:Uplink SWAN} analyzes its uplink performance under the proposed operating protocols, while Section \ref{Section:Downlink SWAN} analyzes the downlink case. Section \ref{Section_Numerical_Results} presents numerical results and provides detailed discussions. Finally, Section \ref{Section_Conclusion} concludes the paper.
\subsubsection*{Notations}
Throughout this paper, scalars, vectors, and matrices are denoted by non-bold, bold lower-case, and bold upper-case letters, respectively. The conjugate and transpose operations are represented by $(\cdot)^{*}$ and $(\cdot)^{\mathsf{T}}$, respectively. The notations $\lvert a\rvert$ and $\lVert \mathbf{a} \rVert$ represent the magnitude of scalar $a$ and the norm of vector $\mathbf{a}$, respectively. The sets $\mathbbmss{C}$ and $\mathbbmss{R}$ denote the complex and real spaces, respectively. The shorthand $[N]$ denotes the set $\{1,\ldots, N\}$. The ceil operator is denoted by $\lceil\cdot\rceil$. The notation ${\mathcal{CN}}(\mu,\sigma^2)$ refers to the circularly symmetric complex Gaussian distribution with mean $\mu$ and variance $\sigma^2$, while ${\mathcal{U}}_{[a,b]}$ denotes the uniform distribution over the interval $[a,b]$. Finally, ${\mathcal{O}}(\cdot)$ denotes big-O notation. 

\section{System Model}\label{Section: System Model}
Consider a communication system in which a BS employs a pinched dielectric waveguide to serve a single-antenna user\footnote{Our previous work demonstrated that orthogonal multiple access schemes (e.g., time-division multiple access and frequency-division multiple access) can approach the capacity limits of PASS \cite{ouyang2025capacity}. Therefore, this paper focuses on a single-antenna typical user served within a single time-frequency resource block without inter-user interference.}. The user is uniformly distributed within a rectangular service region of dimensions $D_x$ and $D_y$ along the $x$- and $y$-axes, respectively. The user location is denoted as ${\mathbf{u}}\triangleq[u_{x},u_{y},0]^{\mathsf{T}}$, as illustrated in {\figurename} {\ref{Figure: PAS_System_Model1}}. The waveguide extends along the $x$-axis to cover the entire horizontal span of the service region. PAs are deployed along the waveguide to radiate signals toward the user or receive signals from the user \cite{suzuki2022pinching}. 
\subsection{Segmented Waveguide Model}
Unlike conventional PASS architectures that employ a single continuous waveguide, we propose the \emph{SWAN} architecture, which adopts a \emph{segmented waveguide} structure composed of multiple short dielectric waveguide segments arranged end-to-end. These segments are \emph{not physically interconnected}. Instead, each segment has its own feed point, through which signals are injected into or extracted from the waveguide and then relayed to the BS via wired links such as optical fiber or high-quality coaxial cables, as shown in {\figurename} {\ref{Figure: PAS_System_Model2}}. Since these wired links incur negligible signal loss compared to the propagation losses within the dielectric segments, they are assumed to be lossless.

The segmented waveguide consists of $M$ segments, each with length $L$, such that $D_x=LM$. When $M=1$, the segmented structure reduces to a conventional single waveguide. Let ${\bm\psi}_{0}^{m}\triangleq[\psi_{0}^{m},\psi_{\rm{w}},d]^{\mathsf{T}}$ denote the location of the feed point of the $m$th segment, where $\psi_{0}^{1}<\psi_{0}^{2}<\ldots<\psi_{0}^{M}$, and $d$ represents the deployment height of the waveguide. For simplicity, the feed point is assumed to be located at the front-left end of each segment. Suppose that $N_m$ PAs are activated along the $m$th segment. The location of the $n$th PA in the $m$th segment (i.e., the $(m,n)$th PA) is denoted by ${\bm\psi}_{n}^{m}\triangleq[\psi_{n}^{m},\psi_{\rm{w}},d]^{\mathsf{T}}$, subject to the following constraints: 
\begin{align}
\psi_{0}^{m}\leq \psi_{n}^{m}\leq \psi_{0}^{m}+L,\lvert\psi_{n'}^{m}-\psi_{n}^{m}\rvert\geq \Delta,\forall n\ne n',
\end{align}
where $\Delta>0$ is the minimum inter-antenna spacing required to mitigate mutual coupling effects \cite{ivrlavc2010toward}.

\begin{figure*}[!t]
\centering
    \subfigure[Segment selection.]
    {
        \includegraphics[height=0.14\textwidth]{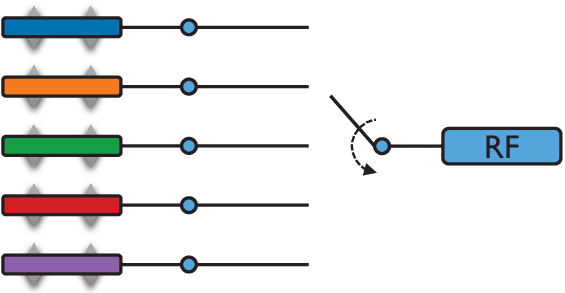}
	   \label{Figure_PAN_Protocol1}
    }
   \subfigure[Segment aggregation.]
    {
        \includegraphics[height=0.14\textwidth]{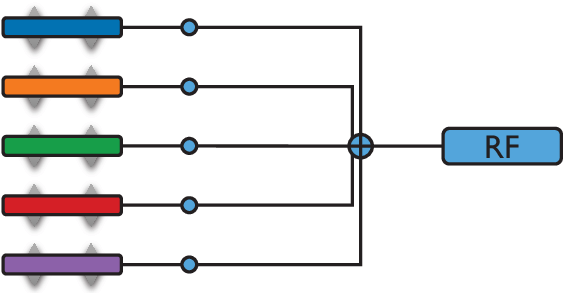}
	   \label{Figure_PAN_Protocol2}
    }
    \subfigure[Segment multiplexing.]
    {
        \includegraphics[height=0.14\textwidth]{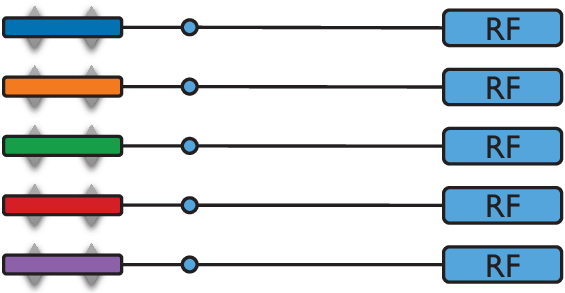}
	   \label{Figure_PAN_Protocol3}
    }
\caption{Illustration of three basic protocols for operating SWANs.}
\label{Figure: PAN_Protocol}
\vspace{-15pt}
\end{figure*}

\subsubsection{Channel Model}
Since PASS is envisioned for high-frequency bands \cite{suzuki2022pinching}, where LoS propagation dominates \cite{ouyang2024primer}, we adopt a free-space LoS channel model to analytically characterize the system performance. Under this model, the spatial channel coefficient between the $(m,n)$th PA and the user is given by \cite{ouyang2024primer}:
\begin{align}
h_{\rm{o}}({\mathbf{u}},{\bm\psi}_{n}^{m})\triangleq
\frac{\eta^{\frac{1}{2}}{\rm{e}}^{-{\rm{j}}k_0\lVert{\mathbf{u}}-{\bm\psi}_{n}^{m}\rVert}}{\lVert{\mathbf{u}}-{\bm\psi}_{n}^{m}\rVert},
\end{align}
where $\eta\triangleq\frac{c^2}{16\pi^2f_{\rm{c}}^2}$, $c$ is the speed of light, $f_{\rm{c}}$ is the carrier frequency, $k_0=\frac{2\pi}{\lambda}$ is the wavenumber, and $\lambda$ is the free-space wavelength. The in-waveguide propagation coefficient between the feed point and the $(m,n)$th PA is modeled as follows \cite{pozar2021microwave}:
\begin{subequations}\label{In_Waveguide_Channel_Model}
\begin{align}
h_{\rm{i}}({\bm\psi}_{n}^{m},{\bm\psi}_{0}^{m})&\triangleq{10^{-\frac{\kappa}{20}\lVert{\bm\psi}_{n}^{m}-{\bm\psi}_{0}^{m}\rVert}}
{\rm{e}}^{-{\rm{j}}\frac{2\pi\lVert{\bm\psi}_{n}^{m}-{\bm\psi}_{0}^{m}\rVert}{\lambda_{\rm{g}}}}
\\&={10^{-\frac{\kappa}{20}\lvert{\psi}_{n}^{m}-{\psi}_{0}^{m}\rvert}}{\rm{e}}^{-{\rm{j}}\frac{2\pi\lvert {\psi}_{n}^{m}-{\psi}_{0}^{m}\rvert}{\lambda_{\rm{g}}} },
\end{align}
\end{subequations}
where $\lambda_{\rm{g}}=\frac{\lambda}{n_{\rm{eff}}}$ is the guided wavelength and $n_{\rm{eff}}$ is the effective refractive index of the dielectric waveguide \cite{pozar2021microwave}. Moreover, $\kappa$ denotes the average attenuation factor along the dielectric waveguide in dB/m \cite{yeh2008essence}. Note that $\kappa=0$ corresponds to the special case of a lossless dielectric and a perfectly conducting surface. Next, we present the signal models for the SWAN-based uplink and downlink channels.
\subsubsection{Uplink SWAN}\label{System_Model_Uplink_SWAN}
In the uplink of the SWAN, the signal received by a PA may re-radiate into free space from another PA along the same waveguide segment as it propagates toward the feed point. This IAR effect makes the signal model mathematically intractable. To simplify the analysis, we assume that only a single PA is activated per segment, i.e., $N_m=1$ for $m\in[M]$. Let $s\sim{\mathcal{CN}}(0,1)$ denote the normalized data symbol transmitted by the user. The received signal at the feed point of the $m$th segment is given by
\begin{align}\label{Uplink_PASS_Basic_Model}
y_{m}^{\rm{ul}}=h_{\rm{i}}({\bm\psi}_{1}^{m},{\bm\psi}_{0}^{m})h_{\rm{o}}({\mathbf{u}},{\bm\psi}_{1}^{m})\sqrt{P}s+n_m^{\rm{ul}},
\end{align}
where $P$ is the transmit power, and $n_m^{\rm{ul}}\sim{\mathcal{CN}}(0,\sigma^2)$ is additive white Gaussian noise (AWGN) with variance $\sigma^2$. 
\subsubsection{Downlink SWAN}
In the downlink of the SWAN, let $x_m$ denote the signal fed into the $m$th waveguide segment. The received signal at the user can be written as follows:
\begin{align}\label{Downlink_PASS_Basic_Model}
y^{\rm{dl}}=\sum_{m=1}^{M}\sum_{n=1}^{N_m}\frac{h_{\rm{i}}({\bm\psi}_{n}^{m},{\bm\psi}_{0}^{m})h_{\rm{o}}({\mathbf{u}},{\bm\psi}_{n}^{m})}{\sqrt{N_m}}x_m+n^{\rm{dl}},
\end{align}
where $n^{\rm{dl}}\sim{\mathcal{CN}}(0,\sigma^2)$ is AWGN. It is assumed that the total transmit power allocated to each waveguide segment is evenly distributed among its $N_m$ active PAs \cite{ding2024flexible,wang2025modeling}. This results in a per-antenna scaling factor of $\frac{1}{\sqrt{N_m}}$ in \eqref{Downlink_PASS_Basic_Model}.
\subsection{Three Basic Operating Protocols}
Referring to {\figurename} {\ref{Figure: PAS_System_Model2}}, the performance of the SWAN depends on the connection mechanism between the feed points and the RF front-end at the BS. Motivated by this observation, we propose three basic operating protocols: \romannumeral1) \emph{segment selection (SS)}, \romannumeral2) \emph{segment aggregation (SA)}, and \romannumeral3) \emph{segment multiplexing (SM)}, as illustrated in {\figurename} {\ref{Figure: PAN_Protocol}}.
\subsubsection{Segment Selection}
For SS, as shown in {\figurename} {\ref{Figure_PAN_Protocol1}}, only a single selected segment is connected to the RF chain at a given time. This protocol is straightforward to implement using a simple switching mechanism and incurs very low hardware complexity. Let ${\bar{m}}\in[M]$ denote the index of the selected segment. The uplink received signal is given by
\begin{align}\label{Uplink_PASS_Basic_Model_SS}
y^{\rm{ul}}=y_{{\bar{m}}}^{\rm{ul}}=h_{\rm{i}}({\bm\psi}_{1}^{{\bar{m}}},{\bm\psi}_{0}^{{\bar{m}}})h_{\rm{o}}({\mathbf{u}},{\bm\psi}_{1}^{{\bar{m}}})
\sqrt{P}s+n_{{\bar{m}}}^{\rm{ul}}.
\end{align}
The downlink signal received at the user is given by
\begin{align}\label{Downlink_PASS_Basic_Model_SS}
y^{\rm{dl}}=\sum_{n=1}^{N_{{\bar{m}}}}\frac{h_{\rm{i}}({\bm\psi}_{n}^{{\bar{m}}},
{\bm\psi}_{0}^{{\bar{m}}})h_{\rm{o}}({\mathbf{u}},{\bm\psi}_{n}^{{\bar{m}}})}{\sqrt{N_{{\bar{m}}}}}x_{{\bar{m}}}+n^{\rm{dl}}.
\end{align}
\subsubsection{Segment Aggregation}
For SA, as illustrated in {\figurename} {\ref{Figure_PAN_Protocol2}}, all $M$ feed points are connected to a single RF chain via a power splitter. In the uplink, signals extracted from all segments are aggregated and forwarded to the RF chain for baseband processing. The aggregated received signal at the BS is given by $y^{\rm{ul}}=\sum_{m=1}^{M}y_{m}^{\rm{ul}}$. It follows from \eqref{Uplink_PASS_Basic_Model} that
\begin{equation}\label{Uplink_PASS_Basic_Model_SA}
y^{\rm{ul}}=\sum_{m=1}^{M}h_{\rm{i}}({\bm\psi}_{1}^{m},{\bm\psi}_{0}^{m})
h_{\rm{o}}({\mathbf{u}}_k,{\bm\psi}_{1}^{m})\sqrt{P}s+{n^{\rm{ul}}},
\end{equation}
where ${n^{\rm{ul}}}\triangleq\sum_{m=1}^{M}n_m^{\rm{ul}}\sim{\mathcal{CN}}(0,M\sigma^2)$ represents the aggregated uplink noise. In the downlink, the transmit signal $x$ is equally split across the $M$ segment, i.e., $x_m=\frac{1}{\sqrt{M}}x$ for $m\in[M]$. The received signal at the user is given by
\begin{align}\label{Downlink_PASS_Basic_Model_SA}
y^{\rm{dl}}=\sum_{m=1}^{M}\sum_{n=1}^{N_m}\frac{h_{\rm{i}}({\bm\psi}_{n}^{m},{\bm\psi}_{0}^{m})h_{\rm{o}}({\mathbf{u}},{\bm\psi}_{n}^{m})}{\sqrt{MN_m}}x+n^{\rm{dl}}.
\end{align}
SA is expected to outperform SS, as all waveguide segments contribute simultaneously, though it requires a more complex RF front-end to enable power splitting and aggregation. 
\subsubsection{Segment Multiplexing}
For SM, as illustrated in {\figurename} {\ref{Figure_PAN_Protocol2}}, each waveguide segment is connected to its own dedicated RF chain. The uplink received signal is given by
\begin{equation}\label{Uplink_PASS_Basic_Model_SM}
{\mathbf{y}}^{\rm{ul}}=[y_{1}^{\rm{ul}},\ldots,y_{M}^{\rm{ul}}]^{\mathsf{T}}={\mathbf{h}}_{\rm{ul}}\sqrt{P}s+{{\mathbf{n}}^{\rm{ul}}},
\end{equation}
where ${\mathbf{h}}_{\rm{ul}}\triangleq[h_{\rm{i}}({\bm\psi}_{1}^{m},{\bm\psi}_{0}^{m})h_{\rm{o}}({\mathbf{u}},{\bm\psi}_{1}^{m})]_{m=1}^{M}\in{\mathbbmss{C}}^{M\times1}$ is the effective uplink channel vector, and ${{\mathbf{n}}^{\rm{ul}}}\triangleq[n_1^{\rm{ul}},\ldots,n_M^{\rm{ul}}]^{\mathsf{T}}\sim{\mathcal{CN}}({\mathbf{0}},\sigma^2{\mathbf{I}}_M)$ is the noise vector. The downlink received signal at the user is expressed as follows:
\begin{equation}\label{Downlink_PASS_Basic_Model_SM}
{{y}}^{\rm{dl}}={\mathbf{h}}_{\rm{dl}}^{\mathsf{T}}{\mathbf{x}}+{{{n}}^{\rm{dl}}},
\end{equation}
where ${\mathbf{x}}\triangleq[x_1,\ldots,x_M]^{\mathsf{T}}\in{\mathbbmss{C}}^{M\times1}$ is the transmit signal vector, and ${\mathbf{h}}_{\rm{dl}}\triangleq[\sum_{n=1}^{N_m}\frac{1}{\sqrt{N_m}}{h_{\rm{i}}({\bm\psi}_{n}^{m},{\bm\psi}_{0}^{m})h_{\rm{o}}({\mathbf{u}},{\bm\psi}_{n}^{m})}]_{m=1}^{M}\in{\mathbbmss{C}}^{M\times1}$ is the downlink effective channel vector. 

\begin{table*}[!t]
\centering
\caption{Summary of the Key Features of the Considered Operating Protocols}
\setlength{\abovecaptionskip}{0pt}
\resizebox{0.95\textwidth}{!}{
\begin{tabular}{|l|l|l|l|l|l|}
\hline
\textbf{Protocol} & \textbf{Architecture}                                  & \textbf{RF Chain} & \textbf{Performance} & \textbf{Implementation Complexity}& \textbf{Optimized Variables} \\ \hline
\textbf{Segment Selection}       & Only one segment connects to the RF chain     & 1        & Lowest      & Very low (switch)  & Antenna positions and activiated segment                \\ \hline
\textbf{Segment Aggregation}       & All segments are aggregated into one RF chain & 1        & Moderate    & Moderate (power splitter)   & Antenna positions        \\ \hline
\textbf{Segment Multiplexing}       & Each segment has its own RF chain             & $M$      & Highest     & High (multi-RF hardware) & Antenna positions and baseband beamforming                     \\ \hline
\end{tabular}}
\label{Table: PASS_Protocol_Comparision}
\vspace{-10pt}
\end{table*}

This setup enables optimal digital processing, such as maximal-ratio combining (MRC) in the uplink and maximal-ratio transmission (MRT) in the downlink, thereby achieving the performance upper bound of the SWAN. However, SM entails significantly higher hardware complexity than SS and SA due to the requirement of multiple RF chains and advanced baseband processing.

Table \ref{Table: PASS_Protocol_Comparision} provides a summary of the key characteristics of the SS, SA, and SM protocols.

\subsubsection{Extension}
\begin{figure}[!t]
\centering
    \subfigure[HSS/A.]
    {
        \includegraphics[height=0.11\textwidth]{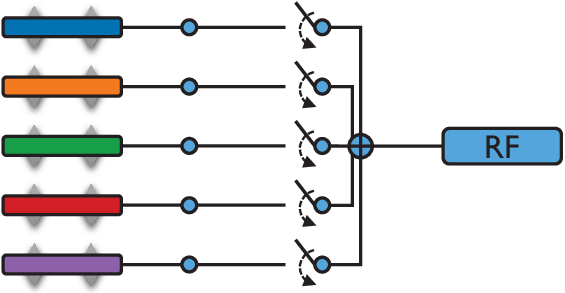}
	   \label{Figure_PAN_Protocol4}
    }
   \subfigure[HSS/M.]
    {
        \includegraphics[height=0.11\textwidth]{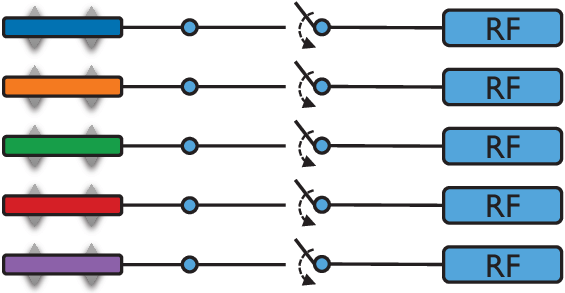}
	   \label{Figure_PAN_Protocol5}
    }
\caption{Illustration of two extensive protocols for SWANs.}
\label{Figure: PAN_Protocol_Ex}
\vspace{-15pt}
\end{figure}

Besides the three basic protocols, new architectures can be designed by combining their key features. For example, a switch can be incorporated into each segment to selectively activate a subset of segments under either SA or SM. These hybrid strategies are referred to as \emph{hybrid segment selection/aggregation (HSS/A)} and \emph{hybrid segment selection/multiplexing (HSS/M)}, respectively, as illustrated in {\figurename} {\ref{Figure: PAN_Protocol_Ex}}. HSS/A provides greater flexibility than both SA and SS: it reduces to SS when only one segment is activated and to SA when all segments are utilized simultaneously. HSS/M reduces hardware complexity compared to SM, since it requires fewer RF chains. While this paper primarily focuses on the three basic protocols in {\figurename} {\ref{Figure: PAN_Protocol}}, the design and analysis of HSS/A and HSS/M constitute promising directions for future research in the SWAN architecture.

\section{Uplink SWAN}\label{Section:Uplink SWAN}
We begin with the uplink case, where PA locations are optimized to maximize the received SNR.
\subsection{Segment Selection}\label{Section: Uplink SWAN: Segment Switching}
\subsubsection{Optimal Antenna Activation}
For the uplink SS protocol, only a single segment and one PA are activated. According to \eqref{Uplink_PASS_Basic_Model_SS}, the maximum received SNR can be written as follows:
\begin{subequations}\label{SNR_SS_Uplink_Definition}
\begin{align}
\gamma_{\rm{SS}}^{\rm{ul}}&\triangleq\max_{{\psi}_{1}^{m}\in[\psi_{0}^{m},\psi_{0}^{m}+L]}
\frac{P\lvert h_{\rm{i}}({\bm\psi}_{1}^{m},{\bm\psi}_{0}^{m})h_{\rm{o}}({\mathbf{u}},{\bm\psi}_{1}^{m})\rvert^2}{\sigma^2}\\
&=\max_{{\psi}_{1}^{m}\in[\psi_{0}^{m},\psi_{0}^{m}+L]}
\frac{P\eta10^{-\frac{\kappa}{10}\lvert{\psi}_{1}^{m}-{\psi}_{0}^{m}\rvert}}{\sigma^2((u_x-{\psi}_{1}^{m})^2+u_y^2+d^2)}.
\end{align}
\end{subequations}
The optimal PA should be activated within the segment closest to the user, whose index $m^{\star}$ is given by
\begin{align}\label{SS_Optimal_Uplink}
m^{\star}=\left\lceil\frac{u_x-{\psi}_{0}^{1}}{L}\right\rceil.
\end{align}  
Substituting this into \eqref{SNR_SS_Uplink_Definition} simplifies the SNR to the following:
\begin{align}
\gamma_{\rm{SS}}^{\rm{ul}}=\max_{{\psi}_{1}^{m^{\star}}\in[\psi_{0}^{m^{\star}},\psi_{0}^{m^{\star}}+L]}
\frac{P\eta10^{-\frac{\kappa}{10}\lvert{\psi}_{1}^{{m^{\star}}}-{\psi}_{0}^{m^{\star}}\rvert}}{\sigma^2((u_x-{\psi}_{1}^{m^{\star}})^2+c_y)},
\end{align}
where $c_y\triangleq (u_y-\psi_{\rm{w}})^2+d^2$. The optimal location of the activated antenna can be obtained as follows \cite[Lemma 1]{xu2025pinching}:
\begin{align}\label{SS_Optimal_Location}
{\psi}_{1}^{m^{\star}}=\left\{\begin{array}{ll}
\psi_{0}^{m^{\star}}             & {c_y\geq\frac{1-(2\alpha u_{m^{\star}}^x-1)^2}{4\alpha^2}}\\
u_x+\frac{-1+\sqrt{1-4\alpha^2c_y}}{2\alpha}           & {\rm{Else}}
\end{array}\right.,
\end{align}
where $\alpha\triangleq\frac{\kappa\ln{10}}{20}$ and $u_{m^{\star}}^x\triangleq u_x-\psi_{0}^{m^{\star}}$. 

It is proven in \cite{xu2025pinching} that, under general PASS configurations, the optimal solution for ${\psi}_{1}^{m^{\star}}$ satisfies 
\begin{align}\label{SS_Approximated_Optimal_Location}
{\psi}_{1}^{m^{\star}}\approx u_x, 
\end{align}
which means that the PA's optimal location nearly aligns with the user's projection onto the waveguide. This placement minimizes the free-space path loss. This approximation will be validated in Section \ref{Section_Numerical_Results} via numerical simulations. Accordingly, the resulting uplink SNR achieved under SS can be approximated as follows:
\begin{align}\label{SNR_SS_Uplink_Optimized}
\gamma_{\rm{SS}}^{\rm{ul}}=
\frac{P}{\sigma^2}\cdot\underbrace{\frac{\eta}{u_y^2+d^2}}_{\text{free-space~path~loss}}\cdot\underbrace{10^{-\frac{\kappa}{10}\lvert u_x-{\psi}_{0}^{m^{\star}}\rvert}}_{\begin{smallmatrix}{\text{in-waveguide}}\\{\text{propagation~loss}}\end{smallmatrix}}.
\end{align}
\subsubsection{Discussion on In-Waveguide Propagation Loss}
A fair baseline for comparison is conventional PASS using a single continuous waveguide and one PA. In this case, the optimal antenna placement also aligns with the user's $x$-coordinate. The corresponding uplink SNR is given by
\begin{align}\label{SNR_CSS_Uplink_Optimized}
\gamma^{\rm{ul}}\triangleq
\frac{P}{\sigma^2}\cdot\underbrace{\frac{\eta}{u_y^2+d^2}}_{\text{free-space~path~loss}}\cdot\underbrace{10^{-\frac{\kappa}{10}\lvert u_x-{\psi}_{0}^{1}\rvert}}_{\begin{smallmatrix}{\text{in-waveguide}}\\{\text{propagation~loss}}\end{smallmatrix}},
\end{align} 
where ${\psi}_{0}^{1}$ is the $x$-coordinate of the feed point of the first waveguide segment. Therefore, it follows that ${\psi}_{0}^{1}\leq {\psi}_{0}^{m^{\star}}$ and $\lvert u_x-{\psi}_{0}^{m^{\star}}\rvert\leq\lvert u_x-{\psi}_{0}^{1}\rvert$.
\vspace{-5pt}
\begin{remark}
Comparing \eqref{SNR_CSS_Uplink_Optimized} and \eqref{SNR_SS_Uplink_Optimized} shows that the SWAN achieves the same free-space path loss as conventional PASS but incurs a smaller in-waveguide propagation loss. This advantage arises because segmentation shortens the effective in-waveguide prorogation distance between the user-aligned PA and the feed point, thereby reducing waveguide attenuation.
\end{remark}
\vspace{-5pt}
Since the user is uniformly distributed within the service region, we have $u_x-{\psi}_{0}^{1}\sim{\mathcal{U}}_{[0,D_x]}$ for conventional PASS and $u_x-{\psi}_{0}^{m^{\star}}\sim{\mathcal{U}}_{[0,L]}$ for SS-based SWAN. Accordingly, the average in-waveguide propagation gain for SS is given by
\begin{subequations}
\begin{align}
A_{\rm{SS}}^{\rm{ul}}&\triangleq \frac{1}{L}\int_{0}^{L}10^{-\frac{\kappa}{10}x}{\rm{d}}x=\frac{1}{L}\frac{1-10^{-\frac{\kappa}{10}L}}{\frac{\kappa}{10}\ln{10}}\\
&=\frac{1-{\rm{e}}^{-2\alpha L}}{2\alpha L}=\frac{1-{\rm{e}}^{-2\alpha \frac{D_x}{M}}}{2\alpha \frac{D_x}{M}},
\end{align}
\end{subequations}
where $D_x=LM$ and $\alpha=\frac{\kappa\ln{10}}{20}$. For conventional PASS, the corresponding average propagation gain is
\begin{align}
A^{\rm{ul}}\triangleq \frac{1}{D_x}\int_{0}^{D_x}10^{-\frac{\kappa}{10}x}{\rm{d}}x
=\frac{1-{\rm{e}}^{-2\alpha D_x}}{2\alpha D_x}.
\end{align}
Next, we examine the behavior of $A_{\rm{SS}}^{\rm{ul}}$ w.r.t. the number of segments $M$.
\vspace{-5pt}
\begin{lemma}\label{Lemma_SS_Derivative}
Given $\alpha$ and $D_x$, the first-order and second-order derivatives of $A_{\rm{SS}}^{\rm{ul}}$ w.r.t. $M$ satisfy
\begin{subequations}
\begin{align}
\frac{\partial A_{\rm{SS}}^{\rm{ul}}}{\partial M}&=\frac{1}{2\alpha D_x}\left(1-\left(1+\frac{2\alpha D_x}{M}\right){\rm{e}}^{-\frac{2\alpha D_x}{M}}\right)>0,\label{Lemma_SS_Derivative_Equation1}\\
\frac{\partial^2A_{\rm{SS}}^{\rm{ul}}}{\partial M^2}&=-\frac{2\alpha D_x}{M^3}{\rm{e}}^{-2\alpha \frac{D_x}{M}}<0.\label{Lemma_SS_Derivative_Equation2}
\end{align}
\end{subequations}
\end{lemma}
\vspace{-5pt} 
\begin{IEEEproof}
These derivatives can be obtained through basic mathematical manipulation. Furthermore, using the inequality ${\rm{e}}^{x}> x+1$ for $x>0$, we have ${\rm{e}}^{\frac{2\alpha D_x}{M}}> \frac{2\alpha D_x}{M}+1$, which yields $\left(1+\frac{2\alpha D_x}{M}\right){\rm{e}}^{-\frac{2\alpha D_x}{M}}<1$. Therefore, $\frac{\partial A_{\rm{SS}}^{\rm{ul}}}{\partial M}>0$. The second derivative is clearly negative.
\end{IEEEproof}
\vspace{-5pt}
\begin{remark}
These results indicate that $A_{\rm{SS}}^{\rm{ul}}$ increases with $M$. Since a larger value of $A_{\rm{SS}}^{\rm{ul}}$ corresponds to lower in-waveguide propagation loss, we conclude that increasing the number of segments $M$ can reduce the in-waveguide propagation loss. This matches intuition: dividing the service region into more segments effectively shortens the average distance between PAs and feed points, thereby reducing signal attenuation within the waveguide.
\end{remark}
\vspace{-5pt}
In conventional PASS, the average in-waveguide prorogation loss scales with the entire side length $D_x$. In contrast, in the SWAN, the effective propagation distance is limited to the segment length $L=\frac{D_x}{M}$. Thus, segmentation changes the scaling law of in-waveguide loss from being proportional to $D_x$ to being proportional to $L$, which significantly mitigates attenuation when $M$ is large. Using the first-order Taylor expansion $\lim_{M\rightarrow\infty}(1-{\rm{e}}^{-2\alpha {D_x}/{M}})\simeq2\alpha \frac{D_x}{M}$, we have
\begin{align}
\lim_{M\rightarrow\infty}A_{\rm{SS}}^{\rm{ul}}=\frac{2\alpha {D_x}/{M}}{2\alpha {D_x}/{M}}=1.
\end{align}
Thus, as $M\rightarrow\infty$, in-waveguide propagation loss becomes negligible, i.e., $A_{\rm{SS}}^{\rm{ul}}\rightarrow1$. This result aligns with intuition: when the number of segments is very large, the feed point of each segment becomes nearly aligned with the projection of the user's location on the waveguide, which minimizes in-waveguide propagation distance.

\begin{figure}[!t]
\centering
\includegraphics[width=0.4\textwidth]{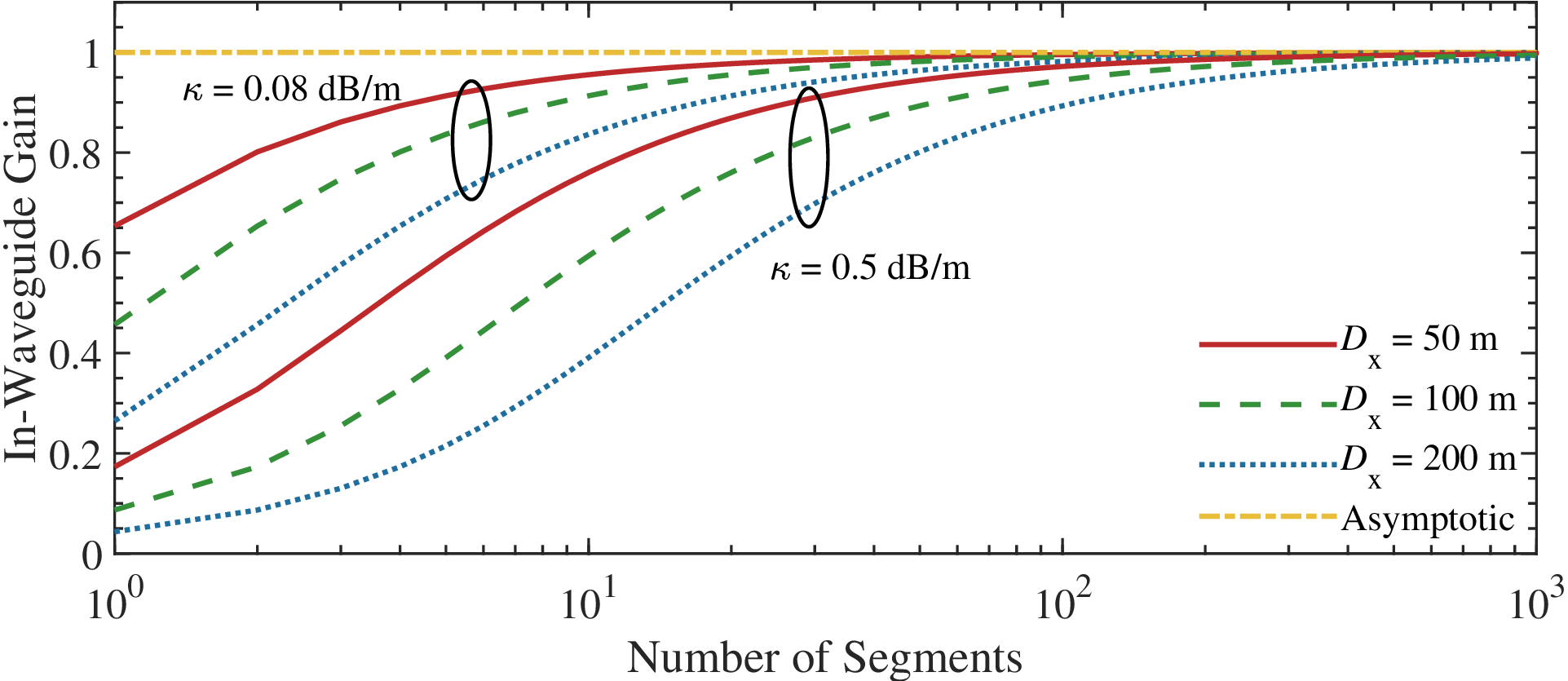}
\caption{Average in-waveguide propagation gain $A_{\rm{SS}}^{\rm{ul}}$ achieved by the SS-based SWAN.}
\label{Figure_In_Waveguide_Gain_Segment}
\vspace{-15pt}
\end{figure}

In practice, however, infinitely many segments are infeasible due to cost and physical constraints. Fortunately, since $A_{\rm{SS}}^{\rm{ul}}$ is concave in $M$ (see \eqref{Lemma_SS_Derivative_Equation2}), the rate of improvement (i.e., reduction in in-waveguide loss) diminishes as $M$ increases, as shown in {\figurename} {\ref{Figure_In_Waveguide_Gain_Segment}}. This property suggests that it is possible to achieve negligible in-waveguide attenuation using a moderate number of segments. For example, consider $D_x=100$ m with attenuation coefficient $\kappa=0.08 $ dB/m ($\alpha=0.0092$ ${\text{m}}^{-1}$) \cite{pozar2021microwave}. To ensure $A_{\rm{SS}}^{\rm{ul}}\geq 0.9$, only $M\geq 9$ segments are required. This shows that negligible propagation loss can be achieved with a modest number of segments in practice.
\subsubsection{Comparison With Conventional PASS}\label{Section: Uplink SWAN: Comparison With Conventional PASS}
Since $A^{\rm{ul}}=A_{\rm{SS}}^{\rm{ul}}$ when $M=1$, it follows that $A_{\rm{SS}}^{\rm{ul}}>A^{\rm{ul}}$ for $M>1$. Comparing the average in-waveguide propagation gains of the segmented and conventional PASS setups yields
\begin{align}
\frac{A_{\rm{SS}}^{\rm{ul}}}{A^{\rm{ul}}}=\frac{{(1-{\rm{e}}^{-2\alpha {D_x}/{M}})M}}{{1-{\rm{e}}^{-2\alpha {D_x}}}}.
\end{align}
According to \eqref{Lemma_SS_Derivative_Equation1}, $\frac{A_{\rm{SS}}^{\rm{ul}}}{A^{\rm{ul}}}$ is monotonically increasing with $M$; that is, employing more segments leads to greater in-waveguide gain improvement. In addition, it can be shown (Appendix \ref{Proof_Derivate_Basic}) that $\frac{\partial}{\partial D_x}\left(\frac{A_{\rm{SS}}^{\rm{ul}}}{A^{\rm{ul}}}\right)>0$, which indicates that this improvement grows with the size of the service region.
\vspace{-5pt}
\begin{remark}\label{Remark_SS_Uplink_SWAN_Gain}
These results imply that the SWAN always experiences lower in-waveguide propagation loss than conventional PASS with a single long waveguide. Moreover, this advantage becomes more pronounced with either a larger service region (i.e., larger $D_x$) or a higher number of segments.
\end{remark}
\vspace{-5pt}
Furthermore, taking the limit as $M\rightarrow\infty$, we obtain
\begin{align}
\lim_{M\rightarrow\infty}\frac{A_{\rm{SS}}^{\rm{ul}}}{A^{\rm{ul}}}=\frac{2\alpha \frac{D_x}{M}M}{{1-{\rm{e}}^{-2\alpha {D_x}}}}=\frac{2\alpha D_x}{{1-{\rm{e}}^{-2\alpha {D_x}}}}>2\alpha D_x,
\end{align}
which shows that, as the number of segments grows, the relative propagation gain of the SWAN over conventional PASS scales at least linearly with the side length $D_{x}$.
\subsection{Segment Aggregation}
We now analyze the performance of the SA-based SWAN. As discussed in Section \ref{Section: Uplink SWAN: Comparison With Conventional PASS}, in-waveguide propagation loss in the SWAN has negligible impact and is therefore omitted. The in-waveguide channel model in \eqref{In_Waveguide_Channel_Model} is simplified as follows:
\begin{align}\label{In_Waveguide_Channel_Model_Simplified}
h_{\rm{i}}({\bm\psi}_{n}^{m},{\bm\psi}_{0}^{m})={\rm{e}}^{-{\rm{j}}\frac{2\pi\lVert{\bm\psi}_{n}^{m}-{\bm\psi}_{0}^{m}\rVert}{\lambda_{\rm{g}}}}={\rm{e}}^{-{\rm{j}}\frac{2\pi\lvert {\psi}_{n}^{m}-{\psi}_{0}^{m}\rvert}{\lambda_{\rm{g}}} }.
\end{align}
Based on \eqref{Uplink_PASS_Basic_Model_SA} and \eqref{In_Waveguide_Channel_Model_Simplified}, the maximum uplink SNR under SA is given by
\begin{subequations}\label{SNR_Uplink_SA_SWAN_Standard}
\begin{align}
\gamma_{\rm{SA}}^{\rm{ul}}&\triangleq\max_{{\bm\psi}_{1}\in{\mathcal{X}}_{\rm{SA}}^{\rm{ul}}}
\frac{P\left\lvert \sum_{m=1}^{M}h_{\rm{i}}({\bm\psi}_{1}^{m},{\bm\psi}_{0}^{m})
h_{\rm{o}}({\mathbf{u}},{\bm\psi}_{1}^{m})\right\rvert^2}{M\sigma^2}\\
&=\max_{{\bm\psi}_{1}\in{\mathcal{X}}_{\rm{SA}}^{\rm{ul}}}
\frac{P\left\lvert \sum\limits_{m=1}^{M}\frac{\eta^{\frac{1}{2}}{\rm{e}}^{-{\rm{j}}k_0(\sqrt{(u_x-{\psi}_{1}^{m})^2+c_y}+n_{\rm{eff}}({\psi}_{1}^{m}-{\psi}_{0}^{m})) }}{\sqrt{(u_x-{\psi}_{1}^{m})^2+c_y}}\right\rvert^2}{M\sigma^2},
\end{align}
\end{subequations}
where ${\bm\psi}_{1}\triangleq[{\psi}_{1}^{1},\ldots,{\psi}_{1}^{M}]^{\mathsf{T}}\in{\mathbbmss{R}}^{M\times1}$ and
\begin{align}
{\mathcal{X}}_{\rm{SA}}^{\rm{ul}}\triangleq\left\{{\bm\psi}_{1}\left\lvert\begin{matrix}{\psi}_{1}^{m}\in[\psi_{0}^{m},\psi_{0}^{m}+L],m\in[M],\\
\lvert{\psi}_{1}^{m}-{\psi}_{1}^{m'}\rvert\geq\Delta,m\ne m'\end{matrix}\right.\right\}.
\end{align}
\subsubsection{Optimal Antenna Placement}\label{SA_Antenna_Optimization_Uplink}
To maximize the received SNR under SA, the positions of the activated PAs must be optimized to achieve constructive signal combining while minimizing free-space path loss \cite{xu2024rate}. For notational convenience, let $M$ be an odd integer.

We first identify the segment that contains the user's projection. Its index is computed as $m^{\star}=\left\lceil\frac{u_x-{\psi}_{0}^{1}}{L}\right\rceil$; see \eqref{SS_Optimal_Uplink}. The antenna in the $m^{\star}$th segment is then placed directly beneath the user's projection, i.e., ${\psi}_{1}^{m^{\star}}=u_x$. Next, consider the $(m^{\star}+1)$th segment. To minimize free-space path loss while satisfying the minimum-spacing constraint $\Delta$, the PA is initially positioned as follows:
\begin{align}\label{SA_Optimization_Update_First}
{\psi}_{1}^{m^{\star}+1}=\max\{{\psi}_{0}^{m^{\star}+1},{\psi}_{1}^{m^{\star}}+\Delta\}\triangleq {\hat{\psi}}_{1}^{m^{\star}+1}.
\end{align} 
We then fine-tune this position to ensure that signals received by the $m^{\star}$th and $(m^{\star}+1)$th antennas experience the same wrapped phase shift, thereby achieving constructive superposition. Specifically, we shift the antenna at ${\hat{\psi}}_{1}^{m^{\star}+1}$ to the right by a distance $\nu_{1}^{m^{\star}+1}>0$, such that
\begin{equation}\label{SA_Slight_Adjustment}
\begin{split}
&({({\hat{\psi}}_{1}^{m^{\star}+1}+\nu_{1}^{m^{\star}+1}-u_x)^2+c_y})^{1/2}+n_{\rm{eff}}({\hat{\psi}}_{1}^{m^{\star}+1}+\nu_{1}^{m^{\star}+1}\\
&-{\psi}_{0}^{m^{\star}+1})
=d_{1}^{m^{\star}+1}+((d_{1}^{m^{\star}}-d_{1}^{m^{\star}+1}))\bmod \lambda)\triangleq \hat{d}_{1}^{m^{\star}+1},
\end{split}
\end{equation}
where $d_{1}^{m^{\star}+1}\triangleq({(\hat{\psi}_{1}^{m^{\star}+1}-u_x)^2+c_y})^{1/2}+n_{\rm{eff}}(\hat{\psi}_{1}^{m^{\star}+1}-{\psi}_{0}^{m^{\star}+1})$ and $d_{1}^{m^{\star}}\triangleq\sqrt{c_y}+n_{\rm{eff}}(u_x-{\psi}_{0}^{m^{\star}})$. The closed-form solution for the shift $\nu_{1}^{m^{\star}+1}$ is given by
\begin{equation}\label{SA_Slight_Adjustment_Solution}
\begin{split}
&\nu_{1}^{m^{\star}+1}=\\&\left\{\begin{array}{ll}
\frac{{\psi}_{0}^{m^{\star}+1}n_{\rm{eff}}^2+\hat{d}_{1}^{m^{\star}+1}n_{\rm{eff}}-u_x-\sqrt{\Delta_{1}^{m^{\star}+1}}}{n_{\rm{eff}}^2-1}-\hat{\psi}_{1}^{m^{\star}+1}             & {n_{\rm{eff}}\ne1}\\
\frac{({\psi}_{0}^{m^{\star}+1}+\hat{d}_{1}^{m^{\star}+1})^2-(u_x^2+c_y)}{2({\psi}_{0}^{m^{\star}+1}+\hat{d}_{1}^{m^{\star}+1}-u_x)}-\hat{\psi}_{1}^{m^{\star}+1}           & {n_{\rm{eff}}=1}
\end{array}\right.,
\end{split}
\end{equation}
where $\Delta_{1}^{m^{\star}+1}\triangleq n_{\rm{eff}}^2(u_x-{\psi}_{0}^{m^{\star}+1})^2-2\hat{d}_{1}^{m^{\star}+1}n_{\rm{eff}}(u_x-{\psi}_{0}^{m^{\star}+1})+c_y(n_{\rm{eff}}^2-1)+(\hat{d}_{1}^{m^{\star}+1})^2$. Since a propagation distance of one wavelength corresponds to a $2\pi$ phase shift, and the left-hand side of \eqref{SA_Slight_Adjustment} increases monotonically with $\nu_{1}^{m^{\star}+1}$, the optimal shift remains within the wavelength scale. This adjustment is significantly smaller than the deployment height $d$, which ensures negligible impact on free-space path loss. 

After obtaining $\nu_{1}^{m^{\star}+1}$, we update the PA position as ${\psi}_{1}^{m^{\star}+1}=\hat{\psi}_{1}^{m^{\star}+1}+\nu_{1}^{m^{\star}+1}$ and set ${\psi}_{1}^{m^{\star}+2}=\max\{{\psi}_{0}^{m^{\star}+2},{\psi}_{1}^{m^{\star}+1}+\Delta\}\triangleq\hat{\psi}_{1}^{m^{\star}+2}$ to maintain the required minimum spacing. By substituting $\hat{\psi}_{1}^{m^{\star}+1}=\hat{\psi}_{1}^{m^{\star}+2}$ into \eqref{SA_Slight_Adjustment} and \eqref{SA_Slight_Adjustment_Solution}, we compute the corresponding shift $\nu_{1}^{m^{\star}+2}$ for the $(m^{\star}+2)$th PA. The same iterative procedure applies for subsequent PAs with $m\geq m_{\star}$, and can be extended symmetrically to antennas with $m< m_{\star}$ \cite{xu2024rate}.
\subsubsection{Discussion on the Maximum SNR}\label{Section: Uplink SA-Based SWAN: Discussion on the Maximum SNR}
The optimization procedure described above enables constructive combination of the received signals \cite{xu2024rate}. As a result, the dual phase shifts induced by signal propagation both inside and outside the waveguide, i.e., $k_0\sqrt{(u_x-{\psi}_{1}^{m})^2+c_y}+k_0n_{\rm{eff}}({\psi}_{1}^{m}-{\psi}_{0}^{m})$, have no impact on the received SNR. Based on \eqref{SNR_Uplink_SA_SWAN_Standard}, we express the uplink SNR as follows: 
\begin{align}
\gamma_{\rm{SA}}^{\rm{ul}}=
\frac{P\eta}{M\sigma^2}\left( \sum\limits_{m=1}^{M}\frac{1}{({(\hat{\psi}_{1}^{m}+\nu_{1}^{m}-u_x)^2+c_y})^{1/2}}\right)^2,
\end{align} 
where $\hat{\psi}_{1}^{m^{\star}}={\psi}_{1}^{m^{\star}}$ and $\nu_{1}^{m^{\star}}=0$. As noted earlier, the shift $\nu_{1}^{m}$ remains within the wavelength scale, and its effect on free-space path loss is negligible. Therefore, the uplink SNR admits the following approximation:
\begin{align}\label{SNR_SA_Uplink_Approximation_First}
\gamma_{\rm{SA}}^{\rm{ul}}\approx
\frac{P\eta}{M\sigma^2}\left( \sum\limits_{m=1}^{M}\frac{1}{({(\hat{\psi}_{1}^{m}-u_x)^2+c_y})^{1/2}}\right)^2.
\end{align}

We now analyze the relationship between the uplink SNR $\gamma_{\rm{SA}}^{\rm{ul}}$ and the number of waveguide segments $M$. For analytical simplicity, suppose the user is located directly beneath the center of the $\left(\frac{M+1}{2}\right)$th waveguide segment. This implies $m^{\star}=\frac{M+1}{2}$ and $u_x=\psi_{m^{\star}}^{0}+\frac{L}{2}$. Moreover, we assume that $L\gg\Delta$, which is a mild condition. Under these assumptions, we have $\hat{\psi}_{m^{\star}+\hat{m}}^{1}=\psi_{m^{\star}+\hat{m}}^{0}$ and $\psi_{m^{\star}-\hat{m}}^{1}=\psi_{m^{\star}-\hat{m}}^{0}+L$ for $\hat{m}=1,\ldots,\frac{M-1}{2}$. Substituting these expressions into \eqref{SNR_SA_Uplink_Approximation_First}, the uplink SNR becomes
\begin{align}
\gamma_{\rm{SA}}^{\rm{ul}}\approx
\frac{P\eta}{M\sigma^2}\left(\frac{1}{\sqrt{c_y}}+\sum\limits_{\hat{m}=1}^{\frac{M-1}{2}}\frac{2}{({(L(\hat{m}-\frac{1}{2}))^2+c_y})^{\frac{1}{2}}}\right)^2.
\end{align}
A tractable expression for $\gamma_{\rm{SA}}^{\rm{ul}}$ can be derived by approximating the summation using the \emph{Euler-Maclaurin formula}.
\vspace{-5pt}
\begin{lemma}\label{Lemma_SA_SWAN_SNR_Uplink}
Given $c_y$, $L$, and $M$, the maximum uplink SNR for the SA-based SWAN can be approximated as follows:
\begin{align}\label{SNR_Approximation_SA_SWAN_SNR_Uplink}
\gamma_{\rm{SA}}^{\rm{ul}}\approx\frac{P\eta\left[\frac{1}{\sqrt{c_y}}+\frac{2\sinh^{-1}\left(\frac{(M-1)L}{2c_y}\right)}{L}
-\frac{\frac{M-1}{24}L^3}{(c_y^2+(\frac{M-1}{2}L)^2)^{\frac{3}{2}}}\right]^2}{M\sigma^2}.
\end{align}
\end{lemma}
\vspace{-5pt}
\begin{IEEEproof}
Please refer to Appendix \ref{Proof_Lemma_SA_SWAN_SNR_Uplink} for more details.
\end{IEEEproof}
For moderate or large $M$, the third term in the brackets of \eqref{SNR_Approximation_SA_SWAN_SNR_Uplink} becomes negligible. Moreover, since $(M-1)L\approx ML=D_x$, the uplink SNR simplifies to the following:
\begin{align}\label{Simplified_SNR_SA_Uplink_Approximation1}
\gamma_{\rm{SA}}^{\rm{ul}}\approx\frac{P\eta}{M\sigma^2}\left(\frac{1}{\sqrt{c_y}}+\frac{2M}{D_x}\sinh^{-1}\left(\frac{D_x}{2c_y}\right)\right)^2.
\end{align}
Inspection of \eqref{Simplified_SNR_SA_Uplink_Approximation1} reveals that when $D_x$ is fixed, $\gamma_{\rm{SA}}^{\rm{ul}}$ is a non-monotonic function of $M$, and it attains a minimum when $M=\frac{D_x}{\sqrt{2c_y}\sinh^{-1}\left(\frac{D_x}{2c_y}\right)}\triangleq M_{\rm{SA}}^{\rm{ul}}$. Specifically, $\gamma_{\rm{SA}}^{\rm{ul}}$ decreases with $M$ when $M\in[1,M_{\rm{SA}}^{\rm{ul}}]$, and increases when $M\in[M_{\rm{SA}}^{\rm{ul}},+\infty)$. This behavior is explained below.
\vspace{-5pt}
\begin{remark}\label{Remark_Uplink_SA_SWAN_Monotone}
When $D_x$ is fixed, most PAs lie far from the user for small $M$. In this regime, increasing $M$ amplifies the total noise power more significantly than it enhances the effective channel gain, which causes the uplink SNR to decrease. For large $M$, more PAs are located near the user, which improves effective channel gain faster than noise growth. Hence, $\gamma_{\rm{SA}}^{\rm{ul}}$ increases with $M$ once the segment density surpasses a certain threshold.
\end{remark}
\vspace{-5pt}
\vspace{-5pt}
\begin{remark}
It can be shown that $M_{\rm{SA}}^{\rm{ul}}$ is a monotonically increasing function of $D_x$, as $\frac{\partial M_{\rm{SA}}^{\rm{ul}}}{\partial D_x}>0$. This observation is consistent with the analysis in Remark \ref{Remark_Uplink_SA_SWAN_Monotone}. As the service region becomes wider (i.e., $D_x$ increases), a greater number of waveguide segments is required to ensure that more PAs are positioned close enough to the user, thereby improving the effective channel gain. 
\end{remark}
\vspace{-5pt}
By examining the asymptotic behavior of \eqref{Simplified_SNR_SA_Uplink_Approximation1}, we obtain
\begin{align}\label{SNR_SA_SWAN_Uplink_High_Segment_Approxiamtion}
\lim_{M\rightarrow\infty}\gamma_{\rm{SA}}^{\rm{ul}}\simeq\frac{4P\eta M}{\sigma^2D_x^2}\left({\sinh^{-1}\left(\frac{D_x}{2c_y}\right)}\right)^2,
\end{align}
which implies $\lim_{M\rightarrow\infty}\gamma_{\rm{SA}}^{\rm{ul}}\simeq{\mathcal{O}}(M)$.
\vspace{-5pt}
\begin{remark}\label{remark_SA_SWAN_Monotine}
Equation \eqref{SNR_SA_SWAN_Uplink_High_Segment_Approxiamtion} demonstrates that when $D_x$ is fixed, the received uplink SNR achieved by the SA-based SWAN scales linearly with the number of waveguide segments $M$. This highlights the benefit of increasing the number of segments to enhance system performance. However, it is important to emphasize that the linear growth $\lim_{M\rightarrow\infty}\gamma_{\rm{SA}}^{\rm{ul}}\simeq{\mathcal{O}}(M)$ does not imply the possibility of achieving arbitrarily large SNR by using an unbounded number of segments. The above derivation assumes the segment length $L\gg\Delta$, which imposes an upper bound on feasible $M$ given a fixed side length $D_x$.
\end{remark}
\vspace{-5pt}
The above discussion assumed a fixed $D_x$. We next consider the case where the segment length $L$ is fixed. In this case, we use the following expression to approximate the SNR:
\begin{align}\label{SNR_Approximation_SA_SWAN_SNR_Uplink2}
\gamma_{\rm{SA}}^{\rm{ul}}\approx\frac{P\eta}{M\sigma^2}\left(\frac{1}{\sqrt{c_y}}+\frac{2}{L}\sinh^{-1}\left(\frac{(M-1)L}{2c_y}\right)\right)^2.
\end{align}
It follows that $\lim_{M\rightarrow\infty}\gamma_{\rm{SA}}^{\rm{ul}}=0$, which suggests that the received SNR does not increase monotonically with the number of segments when the total array aperture $D_x=LM$ is allowed to grow indefinitely. By analyzing the derivative $\frac{\partial \gamma_{\rm{SA}}^{\rm{ul}}}{\partial M}$, it can be shown that the SNR initially increases with $M$, reaches a maximum, and then decreases as $M$ continues to grow; see \cite{ouyang2025array} for further discussion. However, due to the mathematical complexity of \eqref{SNR_Approximation_SA_SWAN_SNR_Uplink2}, a closed-form expression for the optimal $M$ that maximizes $\gamma_{\rm{SA}}^{\rm{ul}}$ is intractable; instead, the transition point can be determined via numerical search.
\vspace{-5pt}
\begin{remark}\label{Remark_Performance_SA_PASS_Optimal_M}
The above result reveals that simply increasing the number of segments $M$ does not guarantee continuous improvement in received SNR. This can be explained as follows: when the length of each segment $L$ is fixed, the aggregate noise power scales linearly with $M$ as $M\sigma^2$, whereas the total received signal power grows only sub-linearly. The sub-linear growth arises because PAs in distant segments contribute weaker signals due to free-space path loss. Therefore, there exists an optimal number of waveguide segment that maximizes the SNR in the SA-based SWAN.
\end{remark}
\vspace{-5pt}
\vspace{-5pt}
\begin{remark}\label{Remark_Performance_SA_PASS_Optimal_HSS}
The above observation suggests that, for a SWAN with fixed-length segments, activating all segments simultaneously may not be beneficial for SA operation. In such cases, HSS/A, as illustrated in {\figurename} {\ref{Figure_PAN_Protocol4}}, offers a practical alternative by adaptively selecting a subset of segments.
\end{remark}
\vspace{-5pt}
\subsubsection{Comparison With Conventional PASS}
We now compare the performance of the SA-based SWAN with that of a conventional uplink PASS employing a single continuous waveguide and a single PA. The case of a conventional PASS with multiple PAs is excluded, since no tractable uplink model exists due to the IAR effect; see Section \ref{System_Model_Uplink_SWAN}. Consequently, the exact performance of a conventional uplink PASS with multiple PAs remains unknown.

Referring to \eqref{SNR_CSS_Uplink_Optimized} and neglecting in-waveguide propagation loss, the uplink SNR for the single-PA conventional PASS is given by $\gamma^{\rm{ul}}=\frac{P\eta}{\sigma^2 c_y}$. Using \eqref{Simplified_SNR_SA_Uplink_Approximation1}, the relative SNR gain of the SA-based SWAN is approximated as follows:
\begin{align}
\frac{\gamma_{\rm{SA}}^{\rm{ul}}}{\gamma^{\rm{ul}}}\approx\frac{1}{M}\left(1+\frac{2\sqrt{c_y}M}{D_x}\sinh^{-1}\left(\frac{D_x}{2c_y}\right)\right)^2.
\end{align}
To ensure that $\frac{\gamma_{\rm{SA}}^{\rm{ul}}}{\gamma^{\rm{ul}}}>1$, it is required that
\begin{align} 
\frac{2\sqrt{c_y}}{D_x}\sinh^{-1}\left(\frac{D_x}{2c_y}\right)>\frac{\sqrt{M}-1}{M},
\end{align} 
which yields the following lower bound on $M$:
\begin{align}
M>\frac{1-\frac{4\sqrt{c_y}\sinh^{-1}\left(\frac{D_x}{2c_y}\right)}{D_x}+\sqrt{1-\frac{8\sqrt{c_y}}{D_x}\sinh^{-1}\left(\frac{D_x}{2c_y}\right)}}
{8\left(\frac{\sqrt{c_y}}{D_x}\sinh^{-1}\left(\frac{D_x}{2c_y}\right)\right)^2}.
\end{align}
Furthermore, taking the limit as $M\rightarrow\infty$, we obtain
\begin{align}\label{Performance_Gain_SA_Uplink_SWAN_PASS}
\lim_{M\rightarrow\infty}\frac{\gamma_{\rm{SA}}^{\rm{ul}}}{\gamma^{\rm{ul}}}\simeq M\left(\frac{2\sqrt{c_y}}{D_x}\sinh^{-1}\left(\frac{D_x}{2c_y}\right)\right)^2,
\end{align}
which indicates that the performance gain of the SA-based SWAN over conventional uplink PASS scales linearly with the number of segments when $M$ is sufficiently large. This performance gain originates from the array gain provided by deploying multiple PAs. Unlike conventional PASS, where multi-PA modeling is hindered by IAR, the segmented SWAN architecture makes multi-PA deployment analytically tractable while avoiding IAR effects.
\subsection{Segment Multiplexing}
We now turn to the SM-based SWAN, where each waveguide segment is connected to a dedicated RF chain. According to \eqref{Uplink_PASS_Basic_Model_SM}, MRC can be applied at baseband to maximize the received SNR, which is given by $\frac{P}{\sigma^2}\lVert{\mathbf{h}}_{\rm{ul}}\rVert^2=\frac{P}{\sigma^2}\sum_{m=1}^{M}\lvert h_{\rm{i}}({\bm\psi}_{1}^{m},{\bm\psi}_{0}^{m})h_{\rm{o}}({\mathbf{u}},{\bm\psi}_{1}^{m})\rvert^2$. By optimizing the PA locations, the maximum achievable SNR under SM can be written as follows:
\begin{align}\label{SNR_Uplink_SM_SWAN_Standard}
\gamma_{\rm{SM}}^{\rm{ul}}=\max_{{\bm\psi}_{1}\in{\mathcal{X}}_{\rm{SA}}^{\rm{ul}}}
\frac{P\eta}{\sigma^2}\sum_{m=1}^{M} \frac{1}{{(u_x-{\psi}_{1}^{m})^2+c_y}}.
\end{align}
\subsubsection{Optimal Antenna Placement}\label{Section: Uplink SWAN: Segment Multiplexing: Optimal Antenna Activation}
Comparing \eqref{SNR_Uplink_SM_SWAN_Standard} with \eqref{SNR_Uplink_SA_SWAN_Standard} reveals that optimizing PA locations in the SM-based SWAN is significantly simpler than in the SA-based SWAN. In the SM case, the objective is to minimize free-space path loss while ensuring the minimum inter-antenna spacing constraint is satisfied. In contrast, SA additionally requires phase alignment across PAs to guarantee constructive superposition, making the optimization problem more complex.

This simplification arises because baseband combining in SM compensates for phase misalignment caused by dual in-waveguide and free-space propagation. Consequently, the optimal PA locations can be determined directly using the placement rule in \eqref{SA_Optimization_Update_First}, without the fine-tuning step described in \eqref{SA_Slight_Adjustment_Solution}. Under this configuration, the resulting uplink SNR of the SM-based SWAN is given by
\begin{align}
\gamma_{\rm{SM}}^{\rm{ul}}=
\frac{P\eta}{\sigma^2}\sum_{m=1}^{M}\frac{1}{{(\hat{\psi}_{1}^{m}-u_x)^2+c_y}},
\end{align}
where $\hat{\psi}_{1}^{m}$ is defined in \eqref{SA_Optimization_Update_First}.
\subsubsection{Discussion on the Maximum SNR}
For analytical simplicity, we assume that $M$ is an odd integer and that the user is located directly beneath the center of the $\left(\frac{M+1}{2}\right)$th waveguide segment. Following a similar derivation to that of \eqref{SNR_Approximation_SA_SWAN_SNR_Uplink}, we approximate $\gamma_{\rm{SM}}^{\rm{ul}}$ as follows:
\begin{align}\label{SNR_Approximation_SM_SWAN_SNR_Uplink}
\gamma_{\rm{SM}}^{\rm{ul}}\approx\frac{P\eta\left[\frac{1}{{c_y}}+\frac{2\tan^{-1}\left(\frac{(M-1)L}{2\sqrt{c_y}}\right)}{L\sqrt{c_y}}
-\frac{\frac{M-1}{12}L^3}{(c_y+(\frac{M-1}{2}L)^2)^{2}}\right]}{\sigma^2}.
\end{align}
Neglecting the third term in the brackets, which is negligible for moderate or large $M$, we further simplify \eqref{SNR_Approximation_SM_SWAN_SNR_Uplink} as follows:
\begin{align}\label{SNR_Approximation_SM_SWAN_SNR_Uplink1}
\gamma_{\rm{SM}}^{\rm{ul}}\approx\frac{P\eta}{\sigma^2}\left(\frac{1}{{c_y}}+\frac{2}{L\sqrt{c_y}}\tan^{-1}\left(\frac{(M-1)L}{2\sqrt{c_y}}\right)\right).
\end{align}
The right-hand side of \eqref{SNR_Approximation_SM_SWAN_SNR_Uplink1} increases monotonically with $M$, which indicates that the use of more waveguide segments always enhance performance. This contrasts with the non-monotonic trend of $\gamma_{\rm{SA}}^{\rm{ul}}$ discussed in Section \ref{Section: Uplink SA-Based SWAN: Discussion on the Maximum SNR}. The key reason for this difference lies in the use of MRC-based optimal combining at baseband in the SM-based architecture. 

Since $(M-1)L<D_x$, we obtain the upper bound:
\begin{align} \label{SNR_Approximation_SM_SWAN_SNR_Uplink_UB}
\gamma_{\rm{SM}}^{\rm{ul}}<\frac{P\eta}{\sigma^2}\left(\frac{1}{{c_y}}+\frac{2}{L\sqrt{c_y}}\tan^{-1}\left(\frac{D_x}{2\sqrt{c_y}}\right)\right). 
\end{align}
This represents a practical performance ceiling for uplink SWAN over a service region of width $D_x$. Importantly, the bound is finite and adheres to the energy conservation law. Taking the limit as $M\rightarrow\infty$, we find $\lim_{M\rightarrow\infty}\tan^{-1}\left(\frac{(M-1)L}{2\sqrt{c_y}}\right)=\frac{\pi}{2}$, which yields
\begin{align}\label{SNR_Approximation_SM_SWAN_SNR_Uplink_UL}
\lim_{M\rightarrow\infty}\gamma_{\rm{SM}}^{\rm{ul}}=\frac{P\eta}{\sigma^2}\left(\frac{1}{{c_y}}+\frac{\pi}{L\sqrt{c_y}}\right).
\end{align}
This expression serves as a general upper bound on the SM-based SWAN, which confirms that the SNR saturates at a finite level in accordance with energy conservation.
\section{Downlink SWAN}\label{Section:Downlink SWAN}
In this section, we extend the previous design and analysis to the downlink SWAN. Unlike the uplink case, it is feasible to deploy \emph{multiple PAs} within each waveguide segment in the downlink to further enhance system performance. When each segment contains only a \emph{single PA}, the downlink SWAN becomes fully \emph{reciprocal (or dual)} to the uplink configuration. In this special case, the derivations and optimization strategies developed for the uplink can be applied directly to the downlink without modification.
\subsection{Segment Selection}\label{Section: Downlink SWAN: Segment Selection}
For the SS-based downlink SWAN, the segment closest to the user is first selected. Its index is given by $m^{\star}=\left\lceil\frac{u_x-{\psi}_{0}^{1}}{L}\right\rceil$; see \eqref{SS_Optimal_Uplink}. We then apply the PA placement method introduced in Section \ref{SA_Antenna_Optimization_Uplink} to optimize the positions of the $N_{m^{\star}}$ PAs within the selected segment. Without loss of generality, we assume that $N_{m^{\star}}$ is an odd integer. 

We first place one antenna directly at the projection of the user's location onto the selected segment, i.e., $\psi_{1}^{m^{\star}}=u_x$. The second antenna is initially placed at $\psi_{1}^{m^{\star}}+\Delta$ and then slightly adjusted, as in \eqref{SA_Slight_Adjustment_Solution}, to align the dual phase shifts. The procedure is repeated to deploy the remaining $\frac{N_{m^{\star}}-1}{2}$ PAs on the right-hand side of $\psi_{1}^{m^{\star}}$, or until the segment boundary is reached. The remaining PAs are then placed on the left side following the same approach. If any computed positions fall outside the segment interval $[\psi_{0}^{m^{\star}},\psi_{0}^{m^{\star}}+L]$, they are reallocated to the opposite side of $\psi_{1}^{m^{\star}}$ within the segment. 

The resulting downlink SNR is expressed as follows:
\begin{align}
\gamma_{\rm{SA}}^{\rm{dl}}=
\frac{P\eta}{N_{m^{\star}}\sigma^2}\left( \sum\limits_{n=1}^{N_{m^{\star}}}\frac{10^{-\frac{\kappa}{20}\lvert \bar{\psi}_{n}^{m^{\star}}-{\psi}_{0}^{m^{\star}}\rvert}}{({(\bar{\psi}_{n}^{m^{\star}}-u_x)^2+c_y})^{1/2}}\right)^2,
\end{align} 
where $\bar{\psi}_{n}^{m^{\star}}\in[\psi_{0}^{m^{\star}},\psi_{0}^{m^{\star}}+L]$ denotes the optimized location of the $n$th PA for $n\in[N_{m^{\star}}]$, and the factor $10^{-\frac{\kappa}{20}\lvert \bar{\psi}_{n}^{m^{\star}}-{\psi}_{0}^{m^{\star}}\rvert}$ accounts for in-waveguide propagation loss. For conventional PASS, a similar PA placement strategy can be applied; however, the PAs are allowed to be distributed over the entire range of the waveguide aperture, i.e., $[\psi_{0}^{1},\psi_{0}^{1}+D_x]$. 

In practice, the inter-antenna spacing $\Delta$ is on the order of the wavelength. If we further assume that $N_{m^{\star}}$ is small, the in-waveguide propagation loss among all PAs in the selected segment is approximately equal:
\begin{align}
10^{-\frac{\kappa}{10}\lvert \bar{\psi}_{n}^{m^{\star}}-{\psi}_{0}^{m^{\star}}\rvert}\approx10^{-\frac{\kappa}{10}\lvert u_x-{\psi}_{0}^{m^{\star}}\rvert},~
n\in[N_{m^{\star}}].
\end{align}
Thus, the same analytical approach as in Section \ref{Section: Uplink SWAN: Segment Switching} applies, and the conclusion remains unchanged: increasing the number of waveguide segments reduces the effective in-waveguide propagation loss. Consequently, the SS-based SWAN achieves superior downlink SNR compared to conventional PASS.

Since the impact of in-waveguide propagation loss is negligible, we omit it in the subsequent analysis. We assume that the user's $x$-coordinate aligns with the center of the selected segment. Under this assumption, and by following the same approach used to derive \eqref{SNR_Approximation_SA_SWAN_SNR_Uplink2}, the downlink SNR of the SS-based SWAN can be approximated as follows:
\begin{align}\label{SNR_Approximation_SS_SWAN_SNR_Downlink2}
\gamma_{\rm{SS}}^{\rm{dl}}\approx\frac{P\eta}{N_{m^{\star}}\sigma^2}\left(\frac{1}{\sqrt{c_y}}+\frac{2}{\Delta}\sinh^{-1}\left(\frac{N_{m^{\star}}-1}{2c_y/\Delta}\right)\right)^2.
\end{align}
It follows that $\gamma_{\rm{SS}}^{\rm{dl}}$ initially increases with $N_{m^{\star}}$, reaches a maximum, and then decreases as $N_{m^{\star}}$ continues to grow. This non-monotonic behavior is analogous to that of $\gamma_{\rm{SA}}^{\rm{ul}}$. Furthermore, as demonstrated in \cite{ouyang2025array}, for typical PASS configurations, the array aperture corresponding to the optimal number of PAs spans tens of meters. 
\vspace{-5pt}
\begin{remark}
As previously discussed, shorter segments yield smaller average in-waveguide propagation loss. In practical designs, such as the prototypes developed by NTT DOCOMO, the length of each waveguide segment is typically less than $1.5$ m \cite{yamamoto2021pinching,reishi2022pinching,junya2024pinching,junya2025pinching}, which is too limited to reach the non-monotonic regime. Therefore, in realistic SS-based SWAN deployments, increasing the number of PAs within a segment can consistently improve system performance.
\end{remark}
\vspace{-5pt}
\subsection{Segment Aggregation}
To achieve additional performance gains, multiple adjacent segments can be activated simultaneously. This motivates the use of SA, which enables multi-segment cooperation for enhanced signal power aggregation. Referring to \eqref{Downlink_PASS_Basic_Model_SA}, the downlink SNR under SA is expressed as follows:
\begin{align}
\gamma_{\rm{SA}}^{\rm{dl}}\triangleq
\max_{{\bm\Psi}\in{\mathcal{X}}_{\rm{SA}}^{\rm{dl}}}
\frac{\left\lvert \sum\limits_{m=1}^{M}\sum\limits_{n=1}^{N_m}\frac{{\rm{e}}^{-{\rm{j}}k_0(\sqrt{(u_x-{\psi}_{n}^{m})^2+c_y}+n_{\rm{eff}}({\psi}_{n}^{m}-{\psi}_{0}^{m})) }}{\sqrt{N_m}\sqrt{(u_x-{\psi}_{n}^{m})^2+c_y}}\right\rvert^2}{M\sigma^2/(P\eta)},
\end{align}
where ${\bm\Psi}\triangleq[{\bm\psi}_{1};\ldots;{\bm\psi}_{N}]\in{\mathbbmss{R}}^{M\times N}$ and
\begin{align}
{\mathcal{X}}_{\rm{SA}}^{\rm{dl}}\triangleq\left\{{\bm\Psi}\left\lvert\begin{matrix}{\psi}_{n}^{m}\in[\psi_{0}^{m},\psi_{0}^{m}+L],m\in[M],n\in[N],\\
\lvert{\psi}_{n}^{m}-{\psi}_{n'}^{m'}\rvert\geq\Delta,(m,n)\ne (m',n')\end{matrix}\right.\right\}.
\end{align}
The optimal locations of the activated PAs can be determined by applying the antenna placement method introduced in Section \ref{SA_Antenna_Optimization_Uplink}, while ensuring phase alignment both within individual segments and across different segments. As discussed in Section \ref{Section: Downlink SWAN: Segment Selection}, the downlink performance is maximized when the effective aperture spanned by the activated PAs reaches a finite optimal size. Thus, for a fixed segment length, activating all available segments is not always beneficial. Instead, the optimal aperture must be identified, and only a subset of segments should be activated. This strategy can be implemented using the HSS/A protocol illustrated in {\figurename} {\ref{Figure_PAN_Protocol4}}.
\subsection{Segment Multiplexing}
The performance of the downlink SWAN can be further improved by employing SM, in which MRT is applied at baseband. According to \eqref{Downlink_PASS_Basic_Model_SM}, the transmit signal for the SM-based SWAN is set as ${\mathbf{x}}=\frac{\sqrt{P}{\mathbf{h}}_{\rm{dl}}^{*}}{\lVert {\mathbf{h}}_{\rm{dl}}\rVert}s$, and the resulting received SNR is expressed as follows:
\begin{align}
\gamma_{\rm{SM}}^{\rm{dl}}\triangleq
\max_{{\bm\Psi}\in{\mathcal{X}}_{\rm{SA}}^{\rm{dl}}}
\frac{\sum\limits_{m=1}^{M}\left\lvert \sum\limits_{n=1}^{N_m}\frac{{\rm{e}}^{-{\rm{j}}k_0(\sqrt{(u_x-{\psi}_{n}^{m})^2+c_y}+n_{\rm{eff}}({\psi}_{n}^{m}-{\psi}_{0}^{m})) }}{\sqrt{N_m}\sqrt{(u_x-{\psi}_{n}^{m})^2+c_y}}\right\rvert^2}{\sigma^2/(P\eta)}.
\end{align}
The optimal placement matrix ${\bm\Psi}$ can be obtained using the method in Section \ref{SA_Antenna_Optimization_Uplink}. Unlike in SA, phase alignment across different segments is not required, since MRT at baseband inherently ensures coherent combining. Only intra-segment phase alignment is necessary. It can be readily shown that the downlink SNR $\gamma_{\rm{SM}}^{\rm{dl}}=\frac{P}{\sigma^2}\lVert{\mathbf{h}}_{\rm{dl}}\rVert^2$ increases with the number of segments $M$. This trend is consistent with the earlier result for the uplink case $\gamma_{\rm{SM}}^{\rm{ul}}$, which confirms that SM achieves the performance upper bound of SWAN.
\section{Numerical Results}\label{Section_Numerical_Results}
We present numerical simulations to validate the analytical results. Unless otherwise specified, the system parameters are set as follows: carrier frequency $f_{\rm{c}} = 28$ GHz, effective refractive index $n_{\rm{eff}}=1.4$, average in-waveguide attenuation factor $\kappa=0.08$ dB/m (equivalently, $\alpha=0.0092$ ${\text{m}}^{-1}$), minimum inter-antenna spacing $\Delta=\frac{\lambda}{2}$, transmit power $P=10$ dBm, and noise power $\sigma^2=-90$ dBm. The user is assumed to be uniformly distributed within a rectangular region centered at the origin, with side lengths $D_x$ and $D_y = 20$ m along the $x$- and $y$-axes, respectively, as illustrated in {\figurename} {\ref{Figure: PAS_System_Model1}}. The waveguide is positioned at a fixed height $d = 3$ m along the $x$-axis with a constant $y$-coordinate of $\psi_{\rm{w}}=0$ m. The $x$-coordinate of the feed point of the first waveguide segment is given by $\psi_{0}^{1}=-\frac{D_x}{2}$. Since the performance gain of PASS over conventional fixed-antenna systems has been extensively demonstrated in the literature \cite{liu2025pinching}, our simulations focus on the comparison between the proposed SWAN and the conventional PASS employing a single long waveguide. All results are averaged over $1000$ independent channel realizations.

\begin{figure}[!t]
\centering
    \subfigure[Average in-waveguide propagation gain vs. $D_x$. $u_y=0$ m.]
    {
        \includegraphics[width=0.4\textwidth]{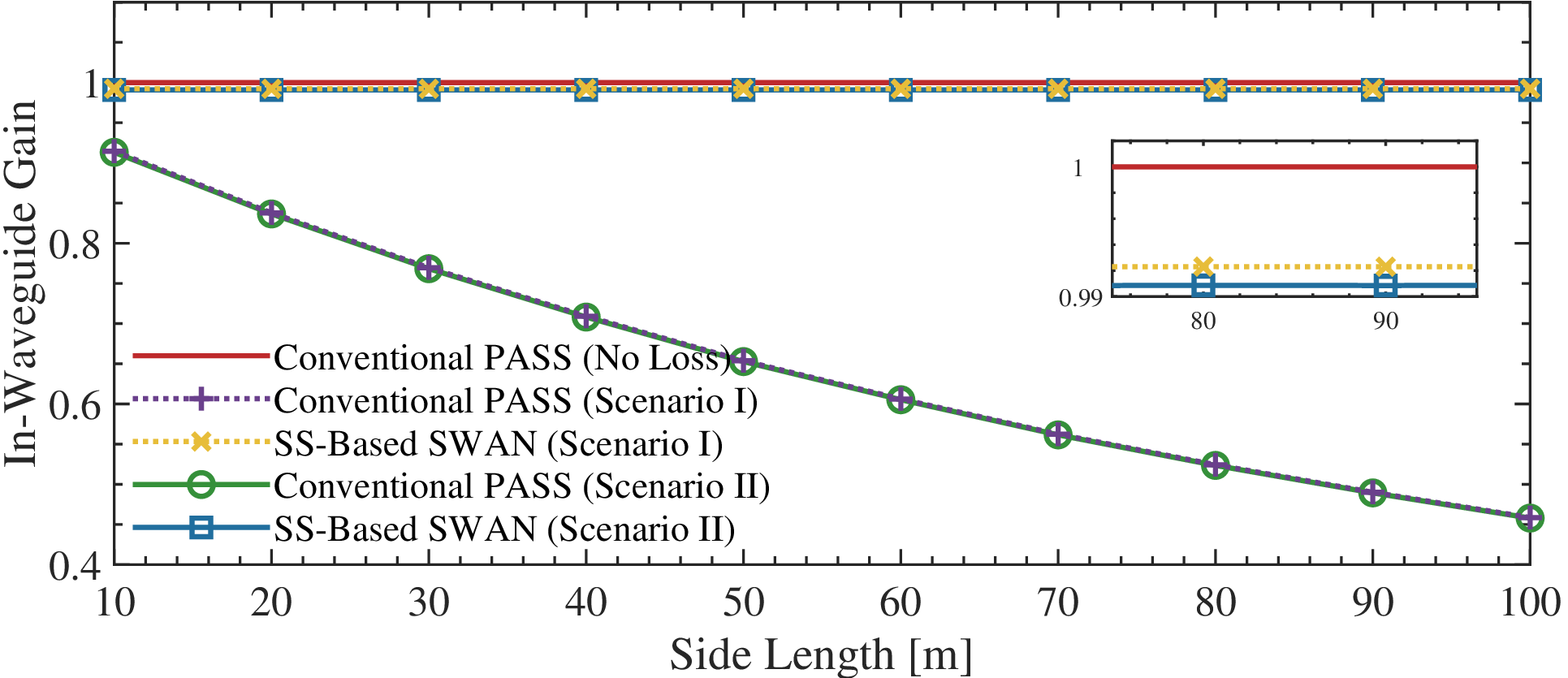}
	   \label{Figure_In_Waveguide_Gain_Side_Length}
    }
   \subfigure[Average user rate vs. $D_x$.]
    {
        \includegraphics[width=0.4\textwidth]{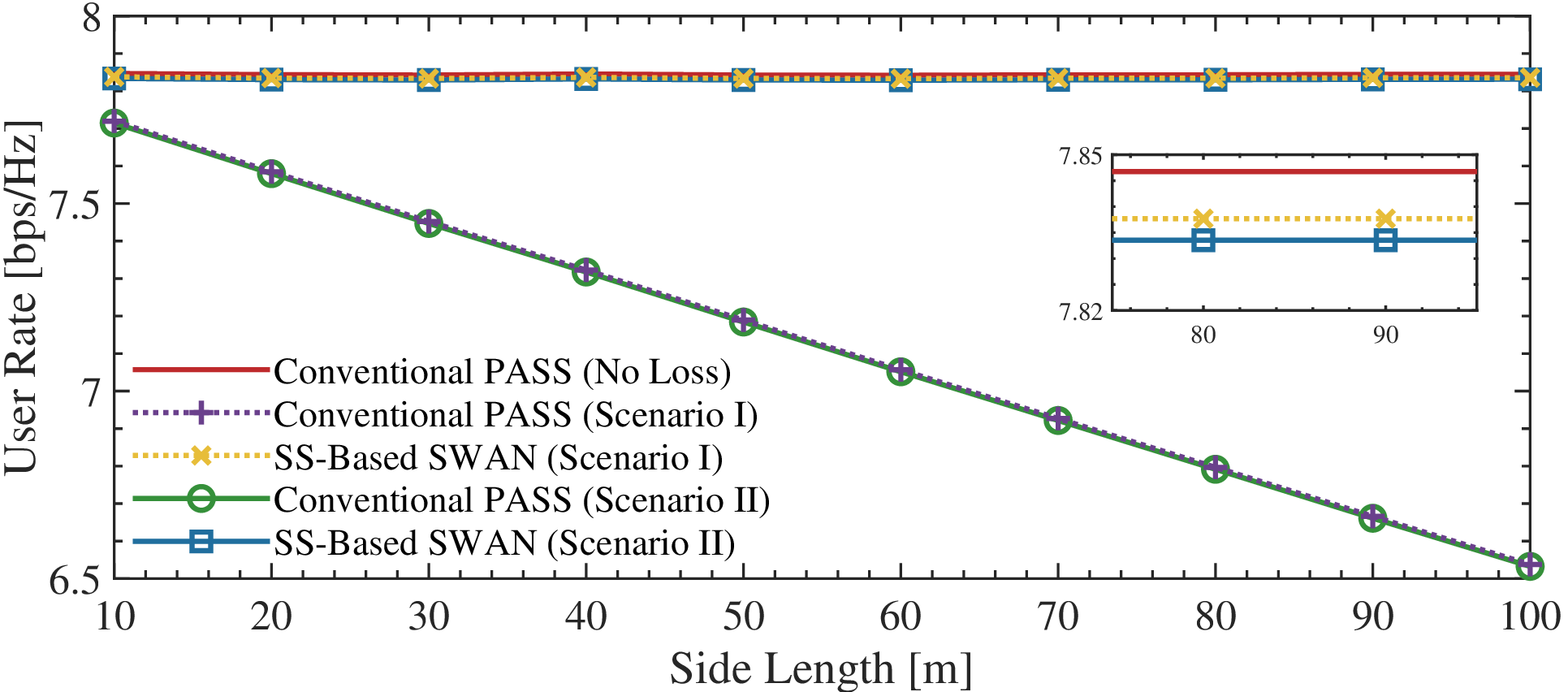}
	   \label{Figure_SS_Uplink_Rate_Side_Length}
    }
\caption{Comparison of the SS-based uplink SWAN and the conventional uplink PASS. $L=1$ m.}
\label{Figure: Uplink SS Comparision}
\vspace{-15pt}
\end{figure}

\subsection{Uplink SWAN}
We first consider the uplink scenario and compare the performance of the proposed SWAN with that of the conventional uplink PASS using a single PA. The case employing multiple PAs in conventional PASS is not considered, as there is currently no tractable or physically consistent model for such systems due to the presence of IRA effects, as discussed in Section \ref{System_Model_Uplink_SWAN}.
\subsubsection{SS-Based SWAN}
We use {\figurename} {\ref{Figure: Uplink SS Comparision}} to validate the theoretical analysis of the SS-based uplink SWAN presented in Section \ref{Section: Uplink SWAN: Segment Switching}. {\figurename} {\ref{Figure_In_Waveguide_Gain_Side_Length}} plots the average in-waveguide propagation gain as a function of the service region width $D_x$, where the user's $y$-coordinate is fixed at $u_y=0$ m and the segment length is set to $L=1$ m. Two PA placement scenarios are considered: \expandafter{\romannumeral1}) Scenario \uppercase\expandafter{\romannumeral1}: optimally placing the PA by \eqref{SS_Optimal_Location}, and \expandafter{\romannumeral2}) Scenario \uppercase\expandafter{\romannumeral2}: placing the PA aligned with the user's projection by \eqref{SS_Approximated_Optimal_Location}. The results show that Scenario \uppercase\expandafter{\romannumeral2} achieves performance close to that of Scenario \uppercase\expandafter{\romannumeral1}, which implies that Scenario \uppercase\expandafter{\romannumeral2} offers a near-optimal solution. Since the analysis in Section \ref{Section: Uplink SWAN: Segment Switching} is derived under Scenario \uppercase\expandafter{\romannumeral2}, this finding suggests that the previous theoretical results represent a performance upper bound for the SS approach. 

{\figurename} {\ref{Figure_In_Waveguide_Gain_Side_Length}} shows that the SS-based SWAN achieves higher in-waveguide propagation gain (equivalently, lower in-waveguide propagation loss) than the conventional uplink PASS. {\figurename} {\ref{Figure_SS_Uplink_Rate_Side_Length}} plots the average achievable user rate as a function of $D_x$ by assuming that the user is uniformly distributed across the rectangular service region. The results demonstrate that SWAN consistently outperforms the conventional PASS in terms of average user rate, primarily due to its reduced in-waveguide propagation loss. {\figurename} {\ref{Figure: Uplink SS Comparision}} shows that the performance of the conventional PASS degrades significantly as the side length $D_x$ increases. This degradation occurs because the conventional PASS relies on a single long waveguide, causing the average distance between the user and the feed point to grow with the size of the service region. As a result, in-waveguide propagation loss becomes more pronounced. In contrast, SWAN deploys multiple short waveguide segments of fixed length. This architecture ensures that the average distance between the user and its corresponding segment remains approximately constant, regardless of the overall width $D_x$ of the region. Consequently, SWAN maintains stable performance across varying values of $D_x$, while the performance of the conventional PASS continues to deteriorate. This widening performance gap, as $D_x$ increases, is consistent with the discussion presented in Remark \ref{Remark_SS_Uplink_SWAN_Gain}. Moreover, because each waveguide segment in SWAN spans only a small portion of the service region, the associated propagation loss remains minimal. The resulting performance approaches that of an ideal, lossless PASS, as shown in {\figurename} {\ref{Figure: Uplink SS Comparision}}. 

\begin{figure}[!t]
\centering
    \subfigure[Received SNR vs. $M$.]
    {
        \includegraphics[width=0.4\textwidth]{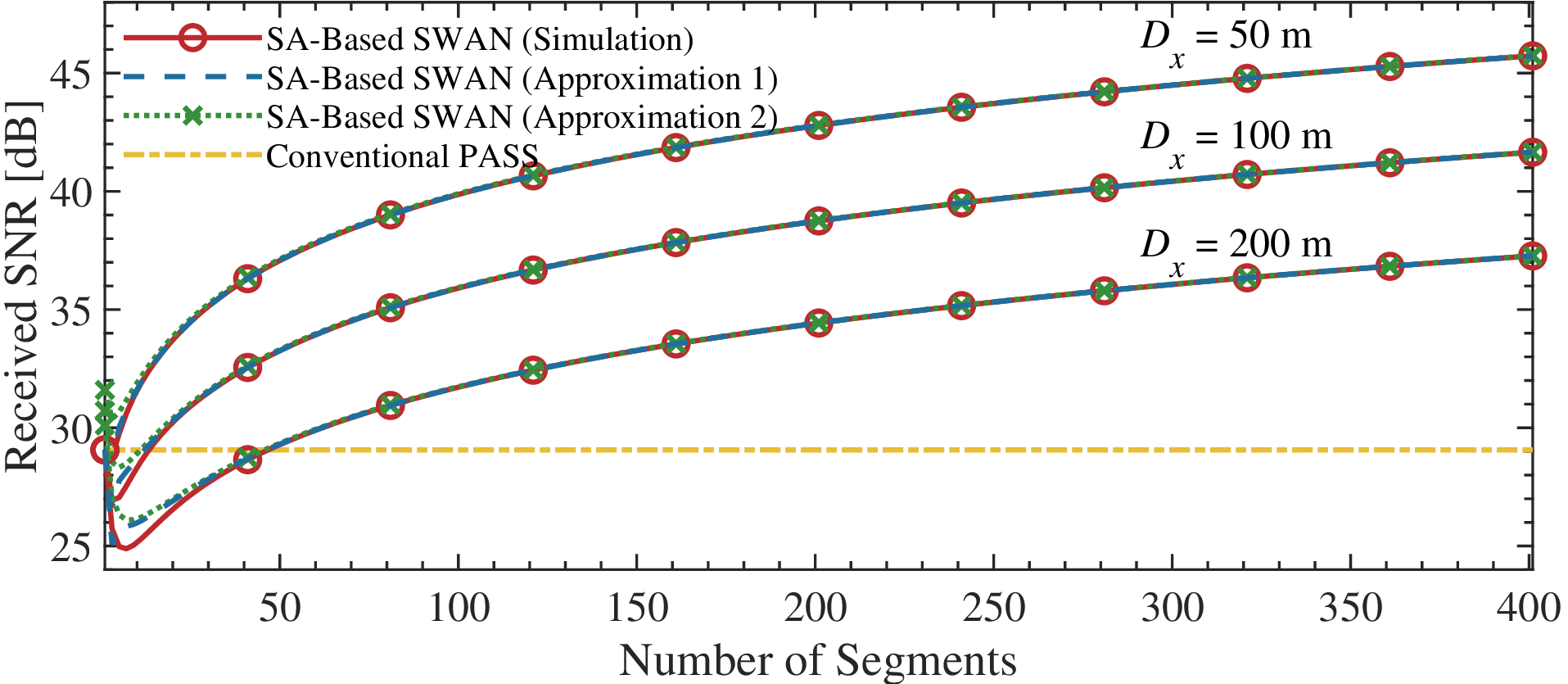}
	   \label{Figure_SA_Uplink_Instantaneous_Received_SNR}
    }
   \subfigure[Received SNR vs. $D_x$.]
    {
        \includegraphics[width=0.4\textwidth]{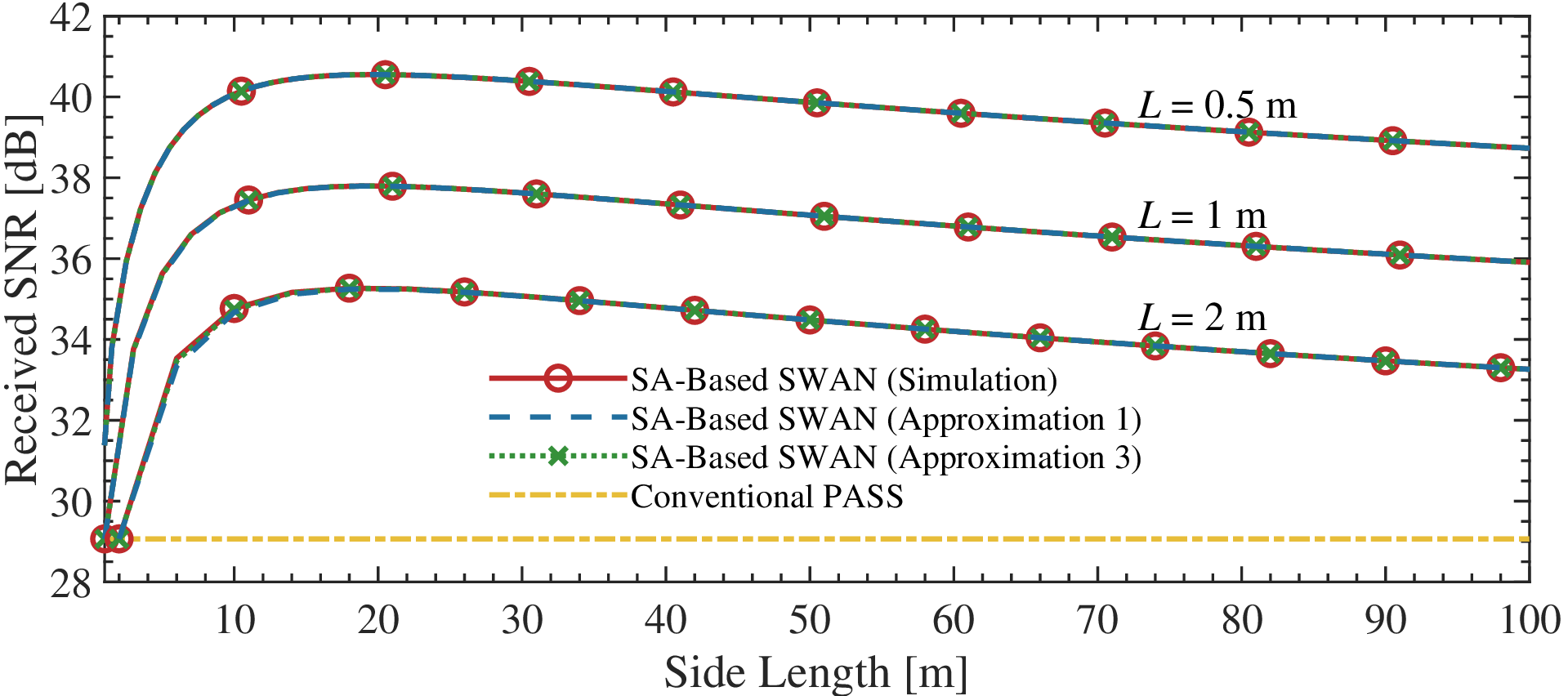}
	   \label{Figure_SA_Uplink_Instantaneous_Received_SNR2}
    }
\caption{SNR comparison of the SA-based uplink SWAN and the conventional PASS. $\kappa=0$ dB/m and $u_x=u_y=0$ m.}
\label{Figure: Uplink SA Comparision}
\vspace{-15pt}
\end{figure}

\subsubsection{SA-Based SWAN}
{\figurename} {\ref{Figure: Uplink SA Comparision}} compares the received SNR of the SA-based SWAN with that of the conventional uplink PASS. Given that in-waveguide propagation loss has negligible impact in SWAN, it is omitted from this comparison. The user's location is fixed at the origin with $u_x=u_y=0$ m. 

{\figurename} {\ref{Figure_SA_Uplink_Instantaneous_Received_SNR}} illustrates the received SNR achieved by the SA-based SWAN as a function of the number of employed waveguide segments $M$ for various values of the side length $D_x$. The simulated SNR is obtained using the method in Section \ref{SA_Antenna_Optimization_Uplink}. The analytical approximations in \eqref{SNR_Approximation_SA_SWAN_SNR_Uplink} (Approximation 1) and \eqref{Simplified_SNR_SA_Uplink_Approximation1} (Approximation 2) closely match the simulation results and accurately capture the SNR trend as a function of $M$. The results show that the SNR of the SA-based SWAN initially decreases with $M$, reaches a minimum, and then rises as $M$ continues to grow. The value of $M$ at which the minimum occurs increases with $D_x$, which supports the discussion in Remark \ref{Remark_Uplink_SA_SWAN_Monotone}. For a fixed $D_x$, increasing $M$ reduces the length of each segment, which allows more PAs to be deployed in closer proximity to the user and enhances the received signal strength. Due to the array gain provided by multiple PAs, the SWAN yields a larger SNR than the conventional PASS, particularly when a larger number of segments is employed. This result aligns with the theoretical prediction in \eqref{Performance_Gain_SA_Uplink_SWAN_PASS}.

{\figurename} {\ref{Figure_SA_Uplink_Instantaneous_Received_SNR2}} plots the received SNR versus the side length $D_x$ for selected values of the segment length $L$. For a fixed $L$, a larger $D_x$ corresponds to a greater number of employed segments $M=\frac{D_x}{L}$. The analytical approximations in \eqref{SNR_Approximation_SA_SWAN_SNR_Uplink} (Approximation 1) and \eqref{SNR_Approximation_SA_SWAN_SNR_Uplink2} (Approximation 3) closely match the simulation results and accurately capture the SNR trend w.r.t. $D_x$. The results show that the SNR of the SA-based SWAN initially increases with $D_x$, reaches a maximum, and then decreases as $D_x$. This confirms the theoretical discussion in Remark \ref{Remark_Performance_SA_PASS_Optimal_M}. The observation further implies that activating all segments is not always beneficial. For example, when $D_x=100$ m, the optimal number of segments spans an aperture of less than $20$ m, as demonstrated in {\figurename} {\ref{Figure_SA_Uplink_Instantaneous_Received_SNR2}}. This highlights the importance of segment selection in practical SWAN deployments and supports the conclusions drawn in Remark \ref{Remark_Performance_SA_PASS_Optimal_HSS}. Moreover, for all considered cases, the SA-based SWAN outperforms the conventional PASS.

\begin{figure}[!t]
\centering
\includegraphics[width=0.4\textwidth]{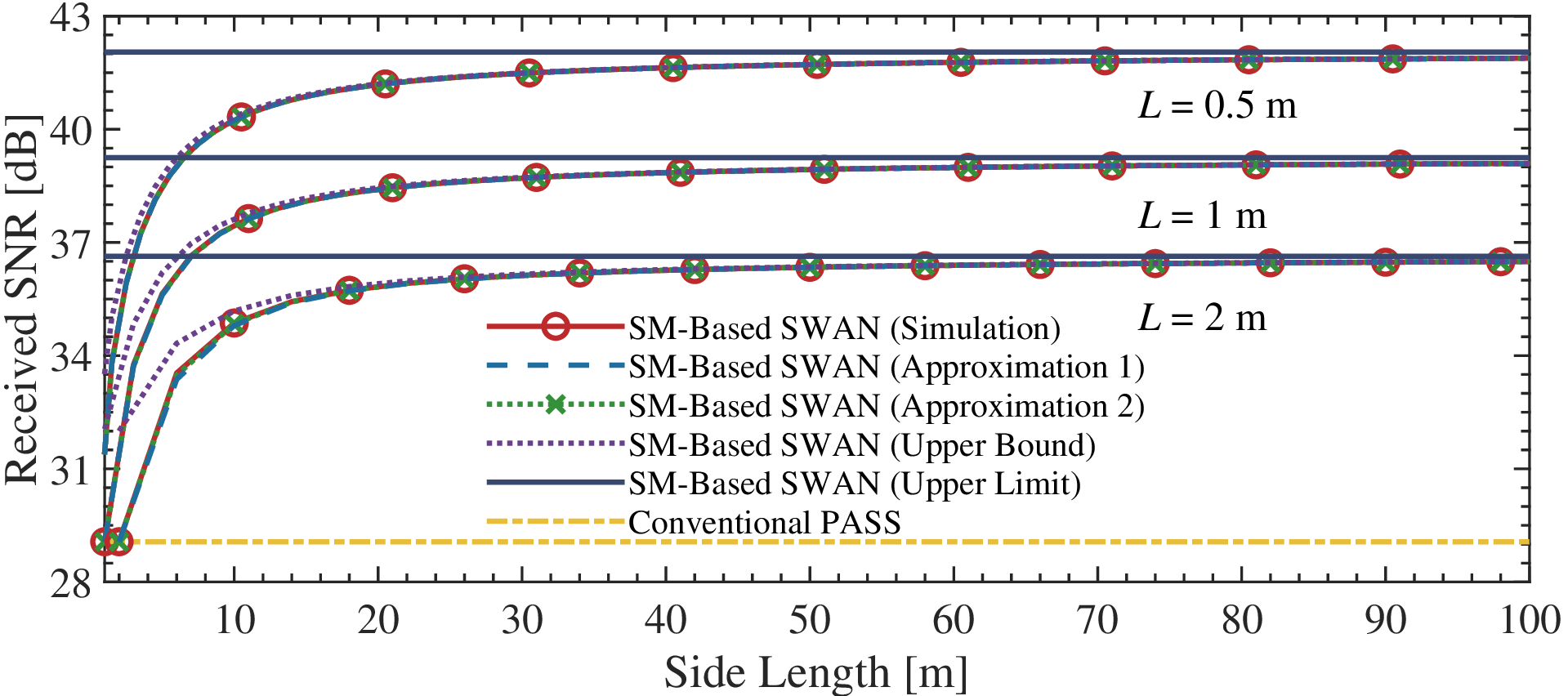}
\caption{SNR comparison of the SM-based uplink SWAN and the conventional PASS. $\kappa=0$ dB/m and $u_x=u_y=0$ m.}
\label{Figure_SM_Uplink_Instantaneous_Received_SNR2}
\vspace{-10pt}
\end{figure}

\subsubsection{SM-Based SWAN}
{\figurename} {\ref{Figure_SM_Uplink_Instantaneous_Received_SNR2}} compares the received SNR of the SM-based SWAN with that of the conventional PASS under the settings $\kappa=0$ dB/m and $u_x=u_y=0$ m. The simulated SNR is obtained using the PA placement method in \ref{Section: Uplink SWAN: Segment Multiplexing: Optimal Antenna Activation}. Its variation w.r.t. $D_x$ is accurately captured by the analytical approximations in \eqref{SNR_Approximation_SM_SWAN_SNR_Uplink} (Approximation 1) and \eqref{SNR_Approximation_SM_SWAN_SNR_Uplink1} (Approximation 2). In addition, the received SNR is upper-bounded by \eqref{SNR_Approximation_SM_SWAN_SNR_Uplink_UB}. For a fixed $L$, as $D_x$ or $M$ increases, the SNR grows monotonically and gradually approaches the upper limit given in \eqref{SNR_Approximation_SM_SWAN_SNR_Uplink_UL}, which is consistent with the energy conservation law. the received SNR achieved by SM increases monotonically with $M$. This behavior contrasts with the results in {\figurename} {\ref{Figure: Uplink SA Comparision}}. The difference arises from the use of MRC-based signal processing in the baseband, enabled by multiple RF chains in the SM-based SWAN.

\begin{figure}[!t]
\centering
\includegraphics[width=0.4\textwidth]{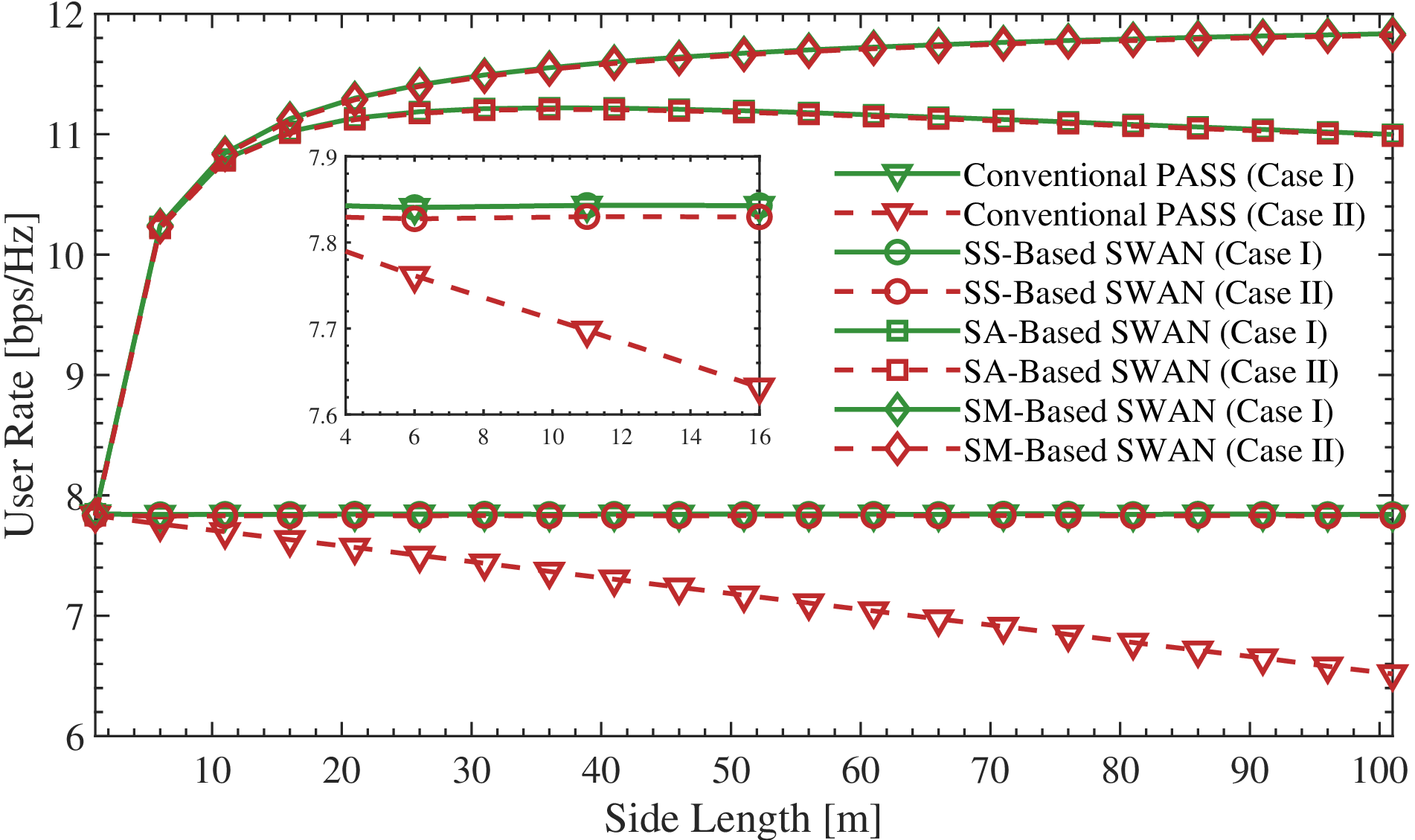}
\caption{Rate comparison of the uplink SWAN and the uplink conventional PASS. $L=1$ m.}
\label{Figure_Uplink_Average_Rate}
\vspace{-15pt}
\end{figure}

\subsubsection{Overall Comparison}
{\figurename} {\ref{Figure_Uplink_Average_Rate}} provides an overall comparison of the SS-, SA-, and SM-based SWAN architectures in terms of the average user rate. For completeness, both the case without in-waveguide propagation loss (Case \uppercase\expandafter{\romannumeral1}) and the case with in-waveguide loss (Case \uppercase\expandafter{\romannumeral2}) are considered. As shown, across all three operation protocols, the SWAN consistently outperforms or matches the performance of the conventional PASS. The performance advantage is more pronounced under Case \uppercase\expandafter{\romannumeral2}, where the effect of in-waveguide propagation loss is included. The results also show that in-waveguide propagation loss has a negligible impact on the performance of SWAN across all three protocols. Furthermore, among the three protocols, SM achieves the highest performance, followed by SA and then SS, which is as expected. Notably, the performance of SA remains close to that of SM, despite its significantly lower hardware complexity. This observation underscores the practicality and superiority of SA in uplink SWAN deployments.
\subsection{Downlink SWAN}

\begin{figure}[!t]
\centering
\includegraphics[width=0.4\textwidth]{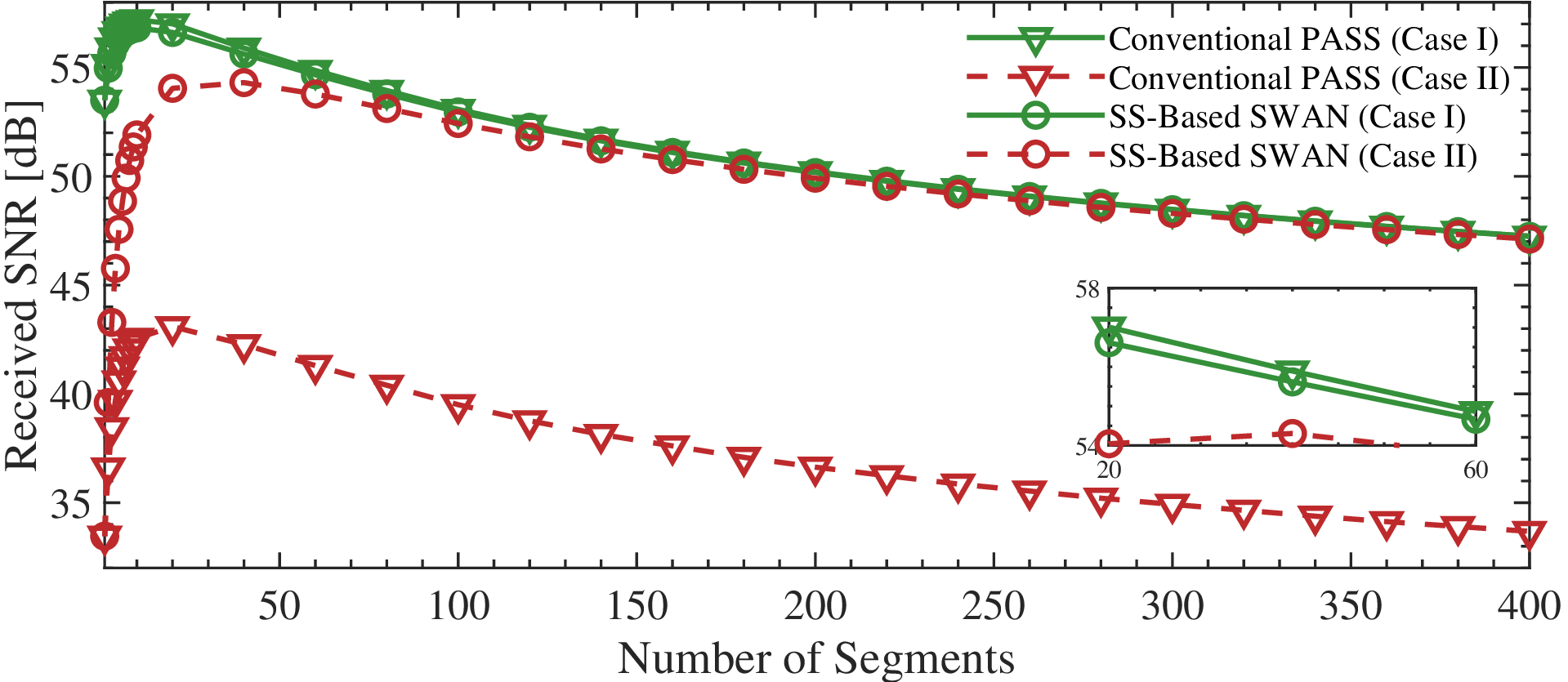}
\caption{SNR comparison of the SS-based downlink SWAN and the conventional PASS. $u_y=0$ m and $D_x=200$ m.}
\label{Figure_SM_Downlink_Received_SNR_Segment}
\vspace{-15pt}
\end{figure}

We next compare the performance of the proposed downlink SWAN with that of the conventional downlink PASS employing multiple PAs. 

{\figurename} {\ref{Figure_SM_Downlink_Received_SNR_Segment}} plots the average received SNR of the SS-based SWAN versus the number of segments $M$ by fixing $u_y=0$ m and $D_x=200$ m. In the SS-based SWAN, PAs are activated on the selected segment as densely as possible, subject to the minimum inter-antenna distance constraint and the segment length constraint. The conventional PASS is configured with the same number of PAs as the SWAN for a fair comparison. Both the lossless case (Case \uppercase\expandafter{\romannumeral1}) and the lossy case (Case \uppercase\expandafter{\romannumeral2}) are evaluated. The results in {\figurename} {\ref{Figure_SM_Downlink_Received_SNR_Segment}} show that the received SNR does not increase monotonically with the number of segments $M$ or the segment length $L=\frac{D_x}{M}$, even when the selected segment is fully populated with PAs. Instead, there exists an optimal segment length, or equivalently, an optimal number of PAs, that maximizes the received SNR. This observation validates the theoretical analysis in \eqref{SNR_Approximation_SS_SWAN_SNR_Downlink2}. 

Furthermore, {\figurename} {\ref{Figure_SM_Downlink_Received_SNR_Segment}} illustrates that, in the absence of in-waveguide propagation loss, the performance of the conventional PASS is slightly better than that of the SS-based SWAN. This is expected, since in the SS-based SWAN, the PAs are restricted to a short segment, whereas in the conventional PASS they may be distributed along the entire side length. However, in practical scenarios where in-waveguide propagation loss is present, the SS-based SWAN significantly outperforms the conventional PASS due to the reduced average distance between the user and the feed point. This performance gain becomes more pronounced with a larger number of segments, or equivalently, shorter segment lengths, which yield a smaller average user-to-feed distance.

\begin{figure}[!t]
\centering
\includegraphics[width=0.4\textwidth]{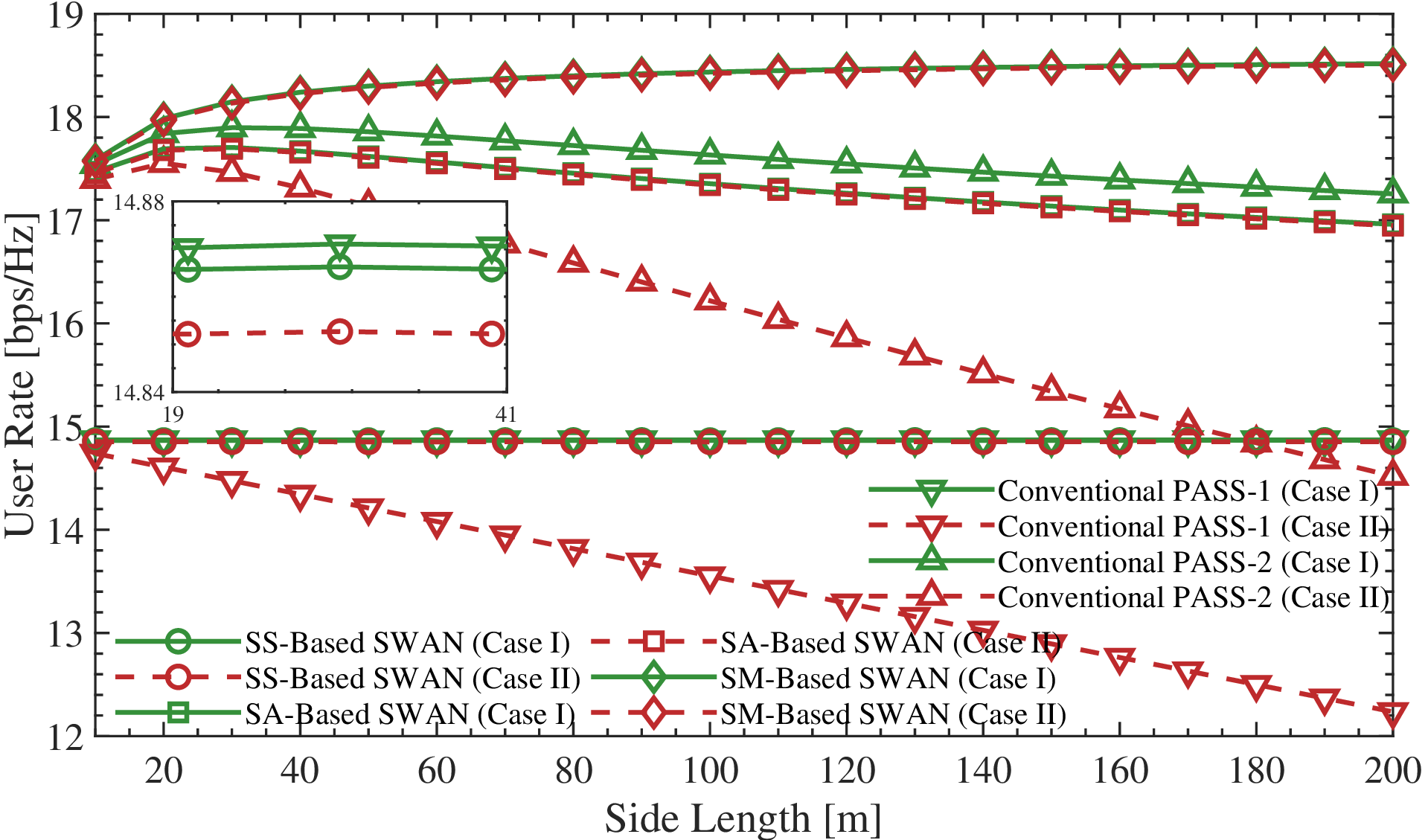}
\caption{Rate comparison of the downlink SWAN and the downlink conventional PASS. $L=1$ m.}
\label{Figure_Downlink_Average_Rate}
\vspace{-15pt}
\end{figure}

{\figurename} {\ref{Figure_Downlink_Average_Rate}} provides an overall comparison of the SS-, SA-, and SM-based SWAN architectures in terms of the average user rate. Under each protocol, PAs are activated on the selected segment or across the entire segmented waveguide as densely as possible, subject to the minimum inter-antenna distance constraint. For reference, two conventional PASS configurations are included. PASS-1 employs the same number of PAs as the SS-based SWAN and serves as its baseline. PASS-2 activates PAs across the entire waveguide as densely as possible, subject to the same minimum spacing, and serves as the baseline for both SA and SM. As expected, SM achieves the highest user rate, followed by SA and then SS. Similar to the results in {\figurename} {\ref{Figure_SM_Downlink_Received_SNR_Segment}}, the conventional PASS slightly outperforms the SA-based SWAN when in-waveguide propagation loss is absent. However, when loss is considered and the side length $D_x$ increases, SA provides a substantial performance advantage over the conventional PASS. For SS, a similar trend is observed: the SS-based SWAN achieves comparable performance to the conventional PASS when in-waveguide propagation loss is absent. However, when this effect is taken into account, the SS-based SWAN significantly outperforms the conventional PASS, since the performance gain from reducing the average user-to-feed distance compensates for the limited flexibility in PA placement due to shorter segment length. This performance advantage becomes more pronounced as $D_x$ grows, because the negative impact of in-waveguide propagation loss intensifies, allowing SS to outperform the conventional PASS.

\section{Conclusion}\label{Section_Conclusion}
This article proposed the SWAN architecture, which employs multiple short waveguide segments with PAs deployed per segment to enable flexible wireless communications. The segmented structure was shown to eliminate the IRA effect, thereby yielding a tractable uplink multi-PA signal model that has not yet been derived in the literature for the conventional multi-PA PASS with a single long waveguide. In addition, the segmented design reduces in-waveguide propagation loss and enhances the overall maintainability of PASS. We proposed three protocols: SS, SA, and SM to facilitate baseband signal processing in SWAN. For each protocol, we proposed PA placement algorithms to maximize the received SNR for both uplink and downlink communications. We further derived closed-form expressions for the SNR under the three protocols and analyzed the corresponding scaling laws w.r.t. the number of segments. These results provided theoretical validation of SWAN's ability to achieve higher SNR than the conventional PASS. Numerical results showed that SWAN reduces both large-scale path loss and in-waveguide propagation loss, which can significantly enhance the SNR compared to the conventional PASS. Among the three protocols, SM achieves the best performance, while SA provides a favorable performance-complexity tradeoff, making it well suited for practical deployment.

This article also reveals new research directions. Both the analysis and simulations showed that an optimal number of segments exists in SA, which motivates joint segment selection and PA placement optimization for hybrid HSS/A. Furthermore, since HSS/M achieves lower computational complexity than SM, future research can investigate joint baseband beamforming, segment selection, and PA placement optimization for HSS/M-based multiuser communications.
\begin{appendix}
\subsection{Proof of $\frac{\partial}{\partial D_x}\left(\frac{A_{\rm{SS}}^{\rm{ul}}}{A_{1}^{\rm{ul}}}\right)>0$}\label{Proof_Derivate_Basic}
Let $f_1(M)\triangleq\frac{1-{\rm{e}}^{-2\alpha D_x/M}}{2\alpha D_x/M}$ and define $R(M,D_x)\triangleq\frac{A_{\rm{SS}}^{\rm{ul}}}{A_{1}^{\rm{ul}}}=\frac{f_1(M)}{f_1(1)}=\frac{M(1-{\rm{e}}^{-2\alpha D_x/M})}{1-{\rm{e}}^{-2\alpha D_x}}$. Let $\beta\triangleq2\alpha D_x$. Because $\beta>0$, by chain rule $\frac{\partial R}{\partial D_x}=\frac{\partial R}{\partial \beta}\frac{\partial \beta}{\partial D_x}=2\alpha\frac{\partial R}{\partial \beta}$, so it suffices to prove $\frac{\partial R}{\partial \beta}>0$. Let $A\triangleq{\rm{e}}^{-\beta/M}\in(0,1)$. Then, ${\rm{e}}^{-\beta}=A^M$ and $R(M,D_x)=\frac{M(1-A)}{1-A^M}$.

Differentiate w.r.t. $\beta$ (noting $\frac{{\rm{d}}A}{{\rm{d}}\beta}=-\frac{1}{M}A$) gives
\begin{align}
\frac{\partial R}{\partial \beta}=\frac{A(1-A^M)-(1-A)MA^M}{(1-A^M)^2}.
\end{align}
The denominator is positive. For the numerator, it can be written as $A(1-f_2(A))$ with $f_2(A)\triangleq A^{M-1}(M-(M-1)A)$. Then for $A\in(0,1)$ and $M\geq1$, it holds that
\begin{align}
f_2'(A)=A^{M-2}(M-1)M(1-A)\geq0,
\end{align}
so $f_2(A)$ is increasing on $A\in(0,1)$. Hence $f_2(A)<f_2(1)=1$, which implies that $A(1-f_2(A))>0$. Consequently, $\frac{\partial R}{\partial \beta}>0$, and since $\beta>0$, we obtain $\frac{\partial}{\partial D_x}\left(\frac{A_{\rm{SS}}^{\rm{ul}}}{A_{1}^{\rm{ul}}}\right)>0$.
\subsection{Proof of Lemma \ref{Lemma_SA_SWAN_SNR_Uplink}}\label{Proof_Lemma_SA_SWAN_SNR_Uplink}
Let $\bar{M}\triangleq\frac{M-1}{2}$ and define $f_3(x)\triangleq\frac{1}{(c_y+x^2)^{1/2}}$. The summation $\sum_{\hat{m}=1}^{\bar{M}}\frac{1}{({(L(\hat{m}-\frac{1}{2}))^2+c_y})^{{1}/{2}}}$ is recognized as the midpoint Riemann sum of $f_3(x)$ over the interval $[0,\bar{M}L]$ with step size $L$. Given that $f_3'(x)=\frac{-x}{(c_y+x^2)^{3/2}}$, we apply the Euler-Maclaurin formula \cite{fornberg2020euler}, which yields
\begin{equation}
\begin{split}
&\sum_{\hat{m}=1}^{\bar{M}}\frac{1}{({(L(\hat{m}-\frac{1}{2}))^2+c_y})^{\frac{1}{2}}}=\frac{1}{L}\int_{0}^{\bar{M}L}f_3(x){\rm{d}}x\\
&+\frac{L^2}{24}(f_3'(\bar{M}L)-f_3'(0))+{\mathcal{O}}(L^4).
\end{split}
\end{equation}
By neglecting higher-order terms and using the integral identity $\int_0^{x} f_3(t){\rm{d}}t=\sinh^{-1}(x/\sqrt{c_y})$, the approximation in \eqref{SNR_Approximation_SA_SWAN_SNR_Uplink} immediately follows. This completes the proof.
\end{appendix}
\bibliographystyle{IEEEtran}
\bibliography{mybib}

\begin{thebibliography}{10}
\providecommand{\url}[1]{#1}
\csname url@samestyle\endcsname
\providecommand{\newblock}{\relax}
\providecommand{\bibinfo}[2]{#2}
\providecommand{\BIBentrySTDinterwordspacing}{\spaceskip=0pt\relax}
\providecommand{\BIBentryALTinterwordstretchfactor}{4}
\providecommand{\BIBentryALTinterwordspacing}{\spaceskip=\fontdimen2\font plus
\BIBentryALTinterwordstretchfactor\fontdimen3\font minus
  \fontdimen4\font\relax}
\providecommand{\BIBforeignlanguage}[2]{{%
\expandafter\ifx\csname l@#1\endcsname\relax
\typeout{** WARNING: IEEEtran.bst: No hyphenation pattern has been}%
\typeout{** loaded for the language `#1'. Using the pattern for}%
\typeout{** the default language instead.}%
\else
\language=\csname l@#1\endcsname
\fi
#2}}
\providecommand{\BIBdecl}{\relax}
\BIBdecl

\bibitem{heath2018foundations}
R.~W. Heath~Jr and A.~Lozano, \emph{Foundations of {MIMO} Communication}.\hskip
  1em plus 0.5em minus 0.4em\relax Cambridge, U.K.: Cambridge Univ. Press,
  2018.

\bibitem{heath2025tri}
R.~W. Heath~Jr, J.~Carlson, N.~V. Deshpande, M.~R. Castellanos, M.~Akrout, and
  C.-B. Chae, ``The tri-hybrid {MIMO} architecture,'' \emph{arXiv preprint
  arXiv:2505.21971}, 2025.

\bibitem{castellanos2023energy}
M.~R. Castellanos, J.~Carlson, and R.~W. Heath, ``Energy-efficient tri-hybrid
  precoding with dynamic metasurface antennas,'' in \emph{Proc. Asilomar Conf.
  Signals, Syst., Comput.}, 2023, pp. 1625--1630.

\bibitem{wong2020fluid}
K.-K. Wong, A.~Shojaeifard, K.-F. Tong, and Y.~Zhang, ``Fluid antenna
  systems,'' \emph{IEEE Trans. Wireless Commun.}, vol.~20, no.~3, pp.
  1950--1962, Mar. 2021.

\bibitem{zhu2024movable}
L.~Zhu, W.~Ma, and R.~Zhang, ``Movable antennas for wireless communication:
  Opportunities and challenges,'' \emph{IEEE Commun. Mag.}, vol.~62, no.~6, pp.
  114--120, Jun. 2024.

\bibitem{suzuki2022pinching}
A.~Fukuda, H.~Yamamoto, H.~Okazaki, Y.~Suzuki, and K.~Kawai, ``Pinching
  antenna: Using a dielectric waveguide as an antenna,'' \emph{NTT DOCOMO
  Technical J.}, vol.~23, no.~3, pp. 5--12, Jan. 2022.

\bibitem{pozar2021microwave}
D.~M. Pozar, \emph{Microwave Engineering: Theory and Techniques}.\hskip 1em
  plus 0.5em minus 0.4em\relax Hoboken, NJ, USA: Wiley, 2021.

\bibitem{yang2025pinching}
Z.~Yang, N.~Wang, Y.~Sun, Z.~Ding, R.~Schober, G.~K. Karagiannidis, V.~W. Wong,
  and O.~A. Dobre, ``Pinching antennas: Principles, applications and
  challenges,'' \emph{arXiv preprint arXiv:2501.10753}, 2025.

\bibitem{liu2025pinching}
Y.~Liu, Z.~Wang, X.~Mu, C.~Ouyang, X.~Xu, and Z.~Ding, ``Pinching antenna
  systems {(PASS)}: Architecture designs, opportunities, and outlook,''
  \emph{IEEE Commun. Mag.}, Accpted to Appear, 2025.

\bibitem{ding2024flexible}
Z.~Ding, R.~Schober, and H.~V. Poor, ``Flexible-antenna systems: A
  pinching-antenna perspective,'' \emph{IEEE Trans. Commun.}, Early Access,
  2025.

\bibitem{tyrovolas2025performance}
D.~Tyrovolas, S.~A. Tegos, P.~D. Diamantoulakis, S.~Ioannidis, C.~K. Liaskos,
  and G.~K. Karagiannidis, ``Performance analysis of pinching-antenna
  systems,'' \emph{IEEE Trans. Cogn. Commun. Netw.}, Early Access, 2025.

\bibitem{ding2025blockage}
Z.~Ding and H.~V. Poor, ``{LoS} blockage in pinching-antenna systems: Curse or
  blessing?'' \emph{IEEE Wireless Commun. Lett.}, Early Access, 2025.

\bibitem{ouyang2025array}
C.~Ouyang, Z.~Wang, Y.~Liu, and Z.~Ding, ``Array gain for pinching-antenna
  systems ({PASS}),'' \emph{IEEE Commun. Lett.}, vol.~29, no.~6, pp.
  1471--1475, Jun. 2025.

\bibitem{xu2024rate}
Y.~Xu, Z.~Ding, and G.~K. Karagiannidis, ``Rate maximization for downlink
  pinching-antenna systems,'' \emph{IEEE Wireless Commun. Lett.}, vol.~14,
  no.~5, pp. 1431--1435, May 2025.

\bibitem{wang2024antenna}
K.~Wang, Z.~Ding, and R.~Schober, ``Antenna activation for {NOMA} assisted
  pinching-antenna systems,'' \emph{IEEE Wireless Commun. Lett.}, vol.~14,
  no.~5, pp. 1526--1530, May 2025.

\bibitem{tegos2024minimum}
S.~A. Tegos, V.~K. Papanikolaou, Z.~Ding, and G.~K. Karagiannidis, ``Minimum
  data rate maximization for uplink pinching-antenna systems,'' \emph{IEEE
  Wireless Commun. Lett.}, vol.~14, no.~5, pp. 1516--1520, May 2025.

\bibitem{zeng2025energy}
M.~Zeng, J.~Wang, G.~Zhou, F.~Fang, and X.~Wang, ``Energy-efficient design for
  downlink pinching-antenna systems with {QoS} guarantee,'' \emph{IEEE Trans.
  Veh. Tech.}, Early Access, 2025.

\bibitem{papanikolaou2025resolving}
V.~K. Papanikolaou, G.~Zhou, B.~Kaziu, A.~Khalili, P.~D. Diamantoulakis, G.~K.
  Karagiannidis, and R.~Schober, ``Resolving the double near-far problem via
  wireless powered pinching-antenna networks,'' \emph{IEEE Wireless Commun.
  Lett.}, Early Access, 2025.

\bibitem{wang2025wireless}
Z.~Wang, C.~Ouyang, Y.~Liu, and A.~Nallanathan, ``Wireless sensing via
  pinching-antenna systems,'' \emph{IEEE Wireless Commun. Lett.}, Early Access,
  2025.

\bibitem{ouyang2025isac}
C.~Ouyang, Z.~Wang, Y.~Zou, Y.~Liu, and Z.~Ding, ``{ISAC} rate region of
  pinching-antenna systems,'' in \emph{IEEE/CIC Int. Conf. Commun. in China
  Workshops}, 2025, pp. 1--6.

\bibitem{ouyang2025rate2}
C.~Ouyang, Z.~Wang, Y.~Liu, and Z.~Ding, ``Rate region of {ISAC} for
  pinching-antenna systems,'' \emph{arXiv preprint arXiv:2505.10179}, 2025.

\bibitem{liu2025pinchingtutorial}
Y.~Liu, H.~Jiang, X.~Xu, Z.~Wang, J.~Guo, C.~Ouyang, X.~Mu, Z.~Ding,
  A.~Nallanathan, G.~K. Karagiannidis \emph{et~al.}, ``Pinching-antenna systems
  ({PASS}): A tutorial,'' \emph{arXiv preprint arXiv:2508.07572}, 2025.

\bibitem{wang2025modeling}
Z.~Wang, C.~Ouyang, X.~Mu, Y.~Liu, and Z.~Ding, ``Modeling and beamforming
  optimization for pinching-antenna systems,'' \emph{arXiv preprint
  arXiv:2502.05917}, 2025.

\bibitem{yamamoto2021pinching}
H.~Yamamoto, A.~Fukuda, H.~Okazaki, and Y.~Suzuki, ``A study on coverage
  extension by a leaky dielectric waveguide with dielectric pieces attached,''
  \emph{IEICE Tech. Rep.}, vol. MW2021, no.~81, pp. 139--144, Nov. 2021.

\bibitem{reishi2022pinching}
R.~Mitani, A.~Fukuda, H.~Yamamoto, H.~Okazaki, Y.~Suzuki, and A.~Nakao,
  ``Research on dielectric waveguides for enhancing spatial resolution in
  beyond {5G} high frequency communications,'' \emph{IEICE Tech. Rep.}, vol.
  NS2022, no. 191, pp. 139--144, Mar. 2023.

\bibitem{junya2024pinching}
J.~Matsudaira, H.~Hamada, A.~Fukuda, F.~Hada, and Y.~Suzuki, ``Evaluation of
  dielectric waveguide connecters,'' \emph{IEICE Tech. Rep.}, vol. MW2024, no.
  138, pp. 75--78, Nov. 2024.

\bibitem{junya2025pinching}
------, ``Evaluation of dielectric waveguide adapters,'' \emph{IEICE Tech.
  Rep.}, vol. ED2024, no.~69, pp. 19--23, Jan. 2025.

\bibitem{ouyang2025capacity}
C.~Ouyang, Z.~Wang, Y.~Liu, and Z.~Ding, ``Capacity characterization of
  pinching-antenna systems,'' \emph{arXiv preprint arXiv:2506.14298}, 2025.

\bibitem{ivrlavc2010toward}
M.~T. Ivrla{\v{c}} and J.~A. Nossek, ``Toward a circuit theory of
  communication,'' \emph{IEEE Trans. Circuits Syst. I, Regular Papers},
  vol.~57, no.~7, pp. 1663--1683, Jul. 2010.

\bibitem{ouyang2024primer}
C.~Ouyang, Z.~Wang, Y.~Chen, X.~Mu, and P.~Zhu, ``A primer on near-field
  communications for next-generation multiple access,'' \emph{Proc. {IEEE}},
  vol. 112, no.~9, pp. 1527--1565, Sep. 2024.

\bibitem{yeh2008essence}
C.~Yeh and F.~I. Shimabukuro, \emph{The Essence of Dielectric
  Waveguides}.\hskip 1em plus 0.5em minus 0.4em\relax New York, NY, USA:
  Springer, 2008.

\bibitem{xu2025pinching}
Y.~Xu, Z.~Ding, R.~Schober, and T.-H. Chang, ``Pinching-antenna systems with
  in-waveguide attenuation: Performance analysis and algorithm design,''
  \emph{arXiv preprint arXiv:2506.23966}, 2025.

\bibitem{fornberg2020euler}
B.~Fornberg, ``{Euler-Maclaurin} expansions without analytic derivatives,''
  \emph{Proc. Royal Soc. London, Series A}, vol. 476, no. 2241, pp. 1--16, Sep.
  2020.

\end{thebibliography}
\end{document}